\documentclass[prb,twocolumn,showpacs,amsmath,amssymb,superscriptaddress]{revtex4-2}

\usepackage{float}
\usepackage{mathtools}
\usepackage{graphicx}
\usepackage{dcolumn}
\usepackage{multirow}
\usepackage{bm}
\usepackage{hyperref}
\hypersetup{pdfnewwindow=true, colorlinks=true, linkcolor=blue, anchorcolor=blue, citecolor=blue, filecolor=blue, menucolor=blue, urlcolor=blue}
\usepackage{color}

\newcolumntype{d}[1]{D{.}{.}{#1}}

\usepackage{pdfpages}
\usepackage{pgffor}
\makeatletter
\AtBeginDocument{\let\LS@rot\@undefined}
\makeatother

\begin{document}

\title{Electric field control of phonon angular momentum in perovskite BaTiO$_3$}

\author{Kevin Moseni}
\email{kevinmoseni@gmail.com}
\affiliation{Materials Science and Engineering, University of California Riverside, Riverside, CA
  92521, USA}
\author{Richard B. Wilson}
\author{Sinisa Coh}
\affiliation{Materials Science and Engineering, University of California Riverside, Riverside, CA
  92521, USA}
\affiliation{Mechanical Engineering, University of California Riverside, Riverside, CA
  92521, USA}

\date{\today}

\begin{abstract}
We find that in BaTiO$_3$ the phonon angular momentum is dominantly pointing in directions perpendicular to the electrical polarization.  Therefore, the external electric field in ferroelectric BaTiO$_3$ does not control only the direction of electrical polarization but also the direction of the phonon angular momentum. This finding opens up the possibility of electric-field control of physical phenomena that depend on phonon angular momentum.  We construct an intuitive model, based on our first-principles calculations, that captures the origin of the relationship between phonon angular momentum and electric polarization.
\end{abstract}

\maketitle

\section{Introduction}\label{sec:Intro} 

In a semi-classical picture phonon modes with angular momentum consist of ions moving about their equilibrium positions either along elliptical or circular paths.~\cite{McLellan,Zhang2014,streib2020difference} Recent experiments have probed such phonon angular momentum in WSe$_2$ with circularly-polarized light.~\cite{Zhang2018} Furthermore, these phonons have been observed to couple to chiral excitons\cite{exciton_coupling,exciton_coupling2} and predicted to couple to topological magnons.\cite{magnon_coupling} Phonon angular momentum is hypothesized to play a key role in the Einstein-de Haas effect,\cite{Zhang2014,nature_EdH} phonon magnetic moment,\cite{Spaldin2018} dynamical multiferroicity\cite{Spaldin2017}, phonon (angular momentum) Hall effects,\cite{PAMHE_pap,PHE_2005} anomalous thermal expansion, \cite{anomalous_therm_exp} negative thermal Hall conductivity,\cite{pseudogap_cuprates_2020} and phonon angular momentum is predicted to be controllable via  temperature gradients\cite{Hamada2018, hamada_thesis}, phonon rotoelectric effect~\cite{hamada2020ph_rotoelectric},  or by straining, doping, and applying a magnetic field in graphene.\cite{graphene_ang_control} Phonons with angular momentum were discussed not only in the context of crystalline phases of matter but also in chiral metamaterials\cite{chiral_meta, 3D_quasicrystal} and in plasmas.\cite{PAM_plasma} 

While some material properties, such as the first-order Raman process, involve phonon excitations at a single point in the Brillouin zone, in this work we focus on physical phenomena that involve phonons at an arbitrary point in the Brillouin zone.  For example, we are interested here in processes such as the ultra-fast electron and phonon dynamics following optical excitation of a material. Since optical excitations can occur at an arbitrary point in the Brillouin zone, the relevant electronic and phonon states also occur at arbitrary points in the Brillouin zone.

Therefore, we need to consider here which group of materials will allow for phonon angular momentum at an arbitrary non-symmetric point of the Brillouin zone.  If we restrict ourselves to the non-magnetic materials, then following Ref.~\onlinecite{classific_arxiv} we see that any non-magnetic material with a broken inversion symmetry will generally have a non-zero phonon angular momentum at a generic non-symmetric point in the Brillouin zone.  Therefore, in this work we focus on non-magnetic materials with broken inversion symmetry.

 Materials with broken inversion symmetry (non-centrosymmetric materials) are either polar or nonpolar.  In the case of a nonpolar material, such as WSe$_2$, the inversion symmetry is broken but the material has no dipole moment.  Therefore, the angular momentum of phonons that arises in a material such as WSe$_2$ will be frozen in the structure, without any way to control it by application of some external perturbation such as an electric field.  On the other hand, the angular momentum of the phonon in a polar material, such as BaTiO$_3$, arises from the displacement of atoms relative to the nonpolar parent crystal structure (in this case cubic). Therefore, depending on the direction and magnitude of the atomic displacements, which can be controlled in ferroelectric BaTiO$_3$ with an electric field, it can be expected that the phonons in the material acquire different directions and magnitudes of phonon angular momentum.  Consequently, physical phenomena that rely on phonon angular momentum could then also, in principle, be controlled with an external electric field. 

In this manuscript, we report on our calculation of the tetragonal and rhombohedral polar phases of BaTiO$_3$. We find that in the tetragonal phase of BaTiO$_3$ the average phonon angular momentum perpendicular to the polar axis is approximately six times larger than along the polar axis.  Due to this anisotropy, the reorientation of the polar axis in tetragonal BaTiO$_3$ could be used to control physical phenomena that depend on the angular momentum of the phonon. We present a simplified model and provide symmetry arguments to understand the origin of the relationship between the direction of the phonon angular momentum and the polar axis. We also computed the phonon angular momentum anisotropy in the rhombohedral phase and found that the anisotropy is three times smaller than in the tetragonal phase.  Although we focus here on BaTiO$_3$ we expect similar effects to occur in other polar perovskites.\cite{STO_model1962,Ghosez1999,Tinte_1999,Seo2013}

In Sec.~\ref{sec:methods} we define some key expressions for phonon angular momentum, then we discuss our first-principles results for the tetragonal phase in Sec.~\ref{sec:results}. In Sec.~\ref{sec:saturation}, we study the phonon angular momentum as we smoothly transform the crystal from polar to nonpolar phase. In Sec.~\ref{sec:anisotropies}, we study the origin of phonon angular momentum anisotropies with a simple model based on our first-principles calculations. In Sec.~\ref{sec:rhomb}, we study the phonon angular momentum in rhombohedral BaTiO$_3$, and then in Sec.~\ref{sec:outlook} we conclude and discuss some possible experiments to observe our predictions. 
 
\section{Methods}\label{sec:methods}

We used the density functional theory approach, as implemented in the computer package \textsc{quantum espresso},\cite{Giannozzi_2009,QE_add_2017} to compute relaxed structures and phonons for bulk BaTiO$_3$. We use the PBEsol exchange-correlation functional.\cite{PBEsol} The ionic potentials are represented with ultrasoft pseudopotentials.\cite{GBRV} We used the kinetic-energy cutoff of 50~Ry for the electron wavefunction and 500~Ry cutoff for the charge density.  We sample the electron Brillouin zone on a $6\times6\times6$ mesh. Our resulting cubic lattice parameter (3.977~\AA) and tetragonal parameters ($a=3.962$~\AA\ and $c=4.058$~\AA) match previously reported theoretical results.\cite{Cooper} We used the linear response method\cite{DFPT} to compute dynamical matrices on the $6\times6\times6$ mesh in the phonon Brillouin zone. We later interpolated this coarse mesh to denser  $30\times30\times30$ $q$-meshes using post-processing tools in \textsc{quantum espresso}. Crystal visualizations were created with the \textsc{vesta}\cite{VESTA} package.

We denote phonon eigenvectors of the dynamical matrix with $\xi_{\bm q \nu}^{i\alpha}$, where $i$ is the atom index in the crystal basis, $\alpha$ is the direction of atomic displacement, while $\bm q$ and $\nu$ are the linear momentum vector and branch index. The matrix of reciprocal space interatomic force constants is denoted as $F_{i j}^{\alpha\beta}(\bm{q})$ using the same conventions.   We obtain the real-space interatomic force-constant matrix by the following Fourier transform,
\begin{equation}\label{eq:elementsofD}
    F_{i j}^{\alpha \beta}(\bm{R}) = \sum_{\bm{q}} e^{-i {\mathbf{q}\cdot \mathbf{R}}}  F^{\alpha\beta}_{ij}(\bm q).
\end{equation}
Therefore $F_{i j}^{\alpha \beta}(\bm{R})$ measures the force induced on the atom $i$ in the direction $\alpha$ due to the displacement of the atom $j$ in the direction $\beta$.  Atoms $i$ and $j$ are generally in different unit cells, separated by a lattice vector $\bm R$.

Given a generic phonon eigenvector $\xi_{\bm q \nu}^{i\alpha}$, one can compute its angular momentum $\bm{l}_{{\bm{q}}\nu}$ following Refs.~\onlinecite{Zhang2014,Zhang2015,McLellan},
\begin{align}
      l^{z}_{{\mathbf{q}}\nu}=
      \sum_i
      2\hbar\left[
      \operatorname{Re}(\xi_{{\mathbf{q}}\nu}^{i x})
      \operatorname{Im}(\xi_{{\mathbf{q}}\nu}^{i y})-
      \operatorname{Re}(\xi_{{\mathbf{q}}\nu}^{i y})
      \operatorname{Im}(\xi_{{\mathbf{q}}\nu }^{i x})\right].
    \label{eq:summed_ion_ang}
\end{align}
Similar expressions hold for the $x$ and $y$ components of the phonon angular momentum. As the expression above is written as a sum over atoms in the unit cell, we can also define for purposes of analysis the contribution of a single atom $i$ to the phonon angular momentum as
\begin{align}
      l^{iz}_{{\mathbf{q}}\nu}=
      2\hbar\left[
      \operatorname{Re}(\xi_{{\mathbf{q}}\nu}^{i x})
      \operatorname{Im}(\xi_{{\mathbf{q}}\nu}^{i y})-
      \operatorname{Re}(\xi_{{\mathbf{q}}\nu}^{i y})
      \operatorname{Im}(\xi_{{\mathbf{q}}\nu }^{i x})\right].
    \label{eq:l}
\end{align}

\section{Results and discussion}\label{sec:results}
\begin{figure}[t]
  \includegraphics[width=3.4in]{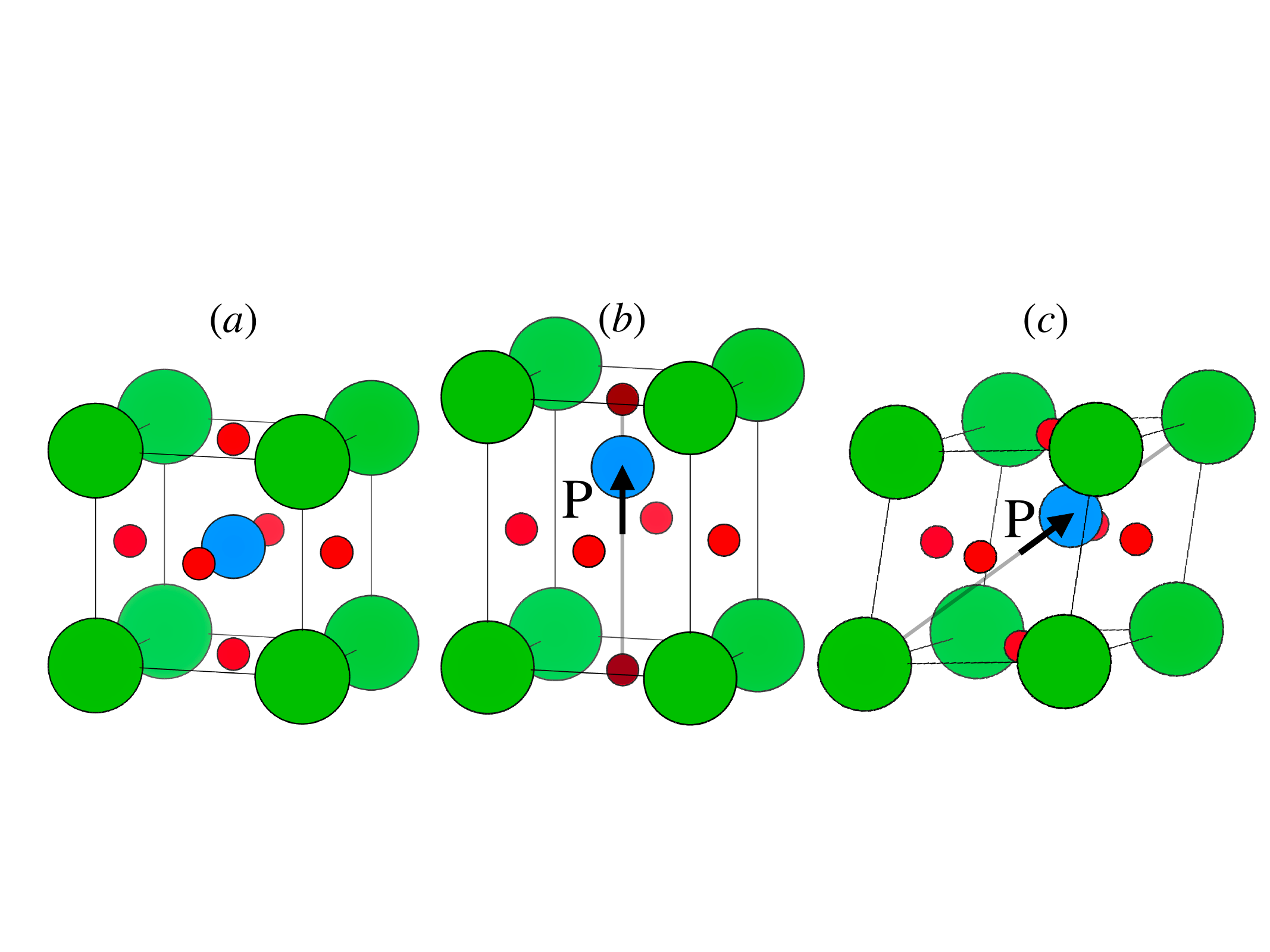}
  \caption{Unit cells of BaTiO$_3$ in (a) cubic, (b) tetragonal, and (c) rhombohedral phases. Ba, Ti, and O atoms are colored green, blue, and red respectively. In (b), we color apical oxygen atoms in darker shade of red. In  (b) and (c) polarization vectors are shown with black arrows.}
  \label{fig:bto_phases}
\end{figure}

BaTiO$_3$ is a prototypical perovskite material with a structural motif consisting of corner--shared Ti--O octahedra. We show the relevant phases of BaTiO$_3$ and their associated polar displacement directions in Figure~\ref{fig:bto_phases}. Above 390~K BaTiO$_3$ can be described as having an average centrosymmetric structure $Pm\bar{3}m$ with a Ti atom in the center of the octahedron, see Fig.~\ref{fig:bto_phases}(a).\cite{BTO_P4mm,kwei} Since this structure has inversion symmetry, phonons in such material will have zero angular momentum.

Between 280~K and 390~K BaTiO$_3$ is in a non-centrosymmetric and polar structure with the space group $P4mm$. Oxygen atoms that are in the same plane as titanium atoms are labeled planar oxygens, while those along the tetragonal axis are labeled apical oxygens.  In the tetragonal phase, polarization develops along one of the pseudocubic axes. In this manuscript we choose the polarization $\bm P$ to point along the $[001]$ direction, as shown in Fig.~\ref{fig:bto_phases}(b). Since inversion symmetry is broken, a generic (non-symmetric) phonon in tetragonal BaTiO$_3$ will now have a non-zero angular momentum $\bm l_{{\bm q} \nu}$. In other words, generically, a phonon will correspond to the elliptical motion of atoms about their equilibrium positions, resulting in a non-zero $\bm l_{{\bm q} \nu}$.  Furthermore, since time-reversal remains a symmetry in BaTiO$_3$ the phonon angular momentum for a given branch at $\bm q$ and $-\bm q$ will be opposite to each other.\footnote{If system has an inversion symmetry but the time-reversal symmetry is broken, then the phonon angular momentum at $\bm q$ and $-\bm q$ has the same sign.} Therefore, the total angular momentum, summed over all $\bm q$ points will be zero, as dictated by the time-reversal symmetry.

At even lower temperatures, between 190~K and 270~K BaTiO$_3$ is in a $Amm2$ structure with polarization pointing along the $[011]$ direction. Lastly, below 15~K BaTiO$_3$ is in a rhombohedral space group $R3m$ with polarization pointing along the pseudocubic $[111]$ direction, as we show in Fig.~\ref{fig:bto_phases}(c).

\subsection{Tetragonal phase}\label{sec:tet}

Now we discuss the calculated phonon angular momentum in the tetragonal phase of BaTiO$_3$.  The distribution of angular momentum ${\bm l}_{{\bm q} \nu}$ is complicated by the fact that all fifteen phonon branches have nonzero ${\bm l}_{{\bm q} \nu}$ at generic $\bm q$. An additional complication is that ${\bm l}_{{\bm q} \nu}$ greatly varies as a function of $\bm q$, especially near the regions where different phonon branches are close in frequency. Therefore, for simplicity, we first analyze the angular momentum of the phonons averaged over branches $\nu$ and wavevectors $\bm q$ over the entire Brillouin zone. An additional reason for taking the average phonon angular momentum is that in this work we are motivated to explore possible physical phenomena that involve electron and phonon dynamics across the Brillouin zone.  For example, in Sec.~\ref{sec:outlook} we suggest a possible way to affect ultra-fast demagnetization in a magnet in close proximity to a ferroelectric such as BaTiO$_3$.  Optical excitation of such a magnet will then generally create electron excitations across the entire Brillouin zone.  Details of the optical excitation will depend on the joint density of states of the metal at the optical excitation energy. Therefore, to get a quantitative measure of the phonon angular momentum in BaTiO$_3$, we decided to simply compute the averages of the phonon angular momenta taken over all $q$-points in the Brillouin zone.

Since in the tetragonal phase polarization $\bm P$ is pointing along the $[001]$ direction we can expect that the averaged phonon angular momentum vector $\bm l$ will have a different magnitude along $\bm P$ and perpendicular to $\bm P$.  For this reason, we define the averages of $\bm l$ perpendicular to ${\bm P}$,
\begin{align}
\label{eq:Lperp}
\big \langle l_{\perp}^{\rm tet} \big \rangle
=
\frac{1}{N_q N_{\nu}} \sum_{\bm q} \sum_{\nu} \sqrt{
\left(
{\bm l}_{{\mathbf{q}}\nu} \cdot {\bm{\hat{n}}}_{100}
\right)^2
+
\left(
{\bm l}_{{\mathbf{q}}\nu} \cdot {\bm{\hat{n}}}_{010}
\right)^2
}
\end{align}
and along $\bm P$,
\begin{align}
\label{eq:Lpar}
\big \langle l_{\parallel}^{\rm tet} \big \rangle
=
\frac{1}{N_q N_{\nu}} \sum_{\bm q} \sum_{\nu} \left| {\bm l}_{{\mathbf{q}}\nu} \cdot {\bm{\hat{n}}}_{001} \right|.
\end{align}
Here, ${\bm{\hat{n}}}$ is the unit vector along the subscripted crystal direction. (Note that averaged quantities defined in Eqs.~\ref{eq:Lperp} and \ref{eq:Lpar} are not affected by the phonon angular momentum at high-symmetry points, lines, or planes, as those parts of the Brillouin zone have zero volume and thus don't contribute to the sum when using a dense enough grid of $\bm q$-points.) We perform averages over positive definite values, as otherwise phonon angular momenta at $\bm q$ and at $-\bm q$ would cancel each other out.  The calculated values of the averaged angular momenta are the following.
\begin{align}
\big \langle l_{\perp}^{\rm tet} \big \rangle
& = 0.154\ \hbar, \notag \\
\big \langle l_{\parallel}^{\rm tet} \big \rangle
& = 0.024 \ \hbar. \notag
\end{align}
Therefore, the angular momentum is about 6 times greater in the plane perpendicular to the polarization $\bm P$ than along $\bm P$. We also note that the magnitude of the average angular momentum is also somewhat large in absolute terms, as the maximum possible phonon angular momentum is $\hbar$. Therefore, the angular momentum of phonons in tetragonal BaTiO$_3$ reaches on average about 15\% of the largest possible value. The estimates of phonon angular momentum in the earlier literature are often reported in systems where the presence of phonon angular momentum is based on the application of an external perturbation.  Furthermore, the phonon angular momentum is usually summed over all phonon branches and q-vectors and weighted by the thermal Bose-Einstein factor.  Therefore, a direct comparison with our result is not straightforward, since the quantities of interest for our work are the sums defined in Eqs.~\ref{eq:Lperp} and \ref{eq:Lpar}.  Nevertheless, we briefly summarize here the phonon angular momentum found in earlier work.  For example, Ref.~\onlinecite{Zhang2014} reports that in CeF$_3$ at an external magnetic field of 6~T the thermally-averaged phonon angular momentum is around $0.02~\hbar$ per one unit cell.  Similarly, Ref.~\onlinecite{hamada2020ph_rotoelectric} reports that the average phonon angular momentum in Cr$_2$O$_3$ under 10~V/mm electric field is around $10^{-8}~\hbar$ per unit cell.  Other works deal with situations such as high-symmetry points in the Brillouin zone, where the phonon eigenvector is fully circularly polarized with the phonon angular momentum of $\pm \hbar$.\cite{Zhang2015,Hamada2018,Spaldin2018}

\begin{figure}[t]
  \includegraphics{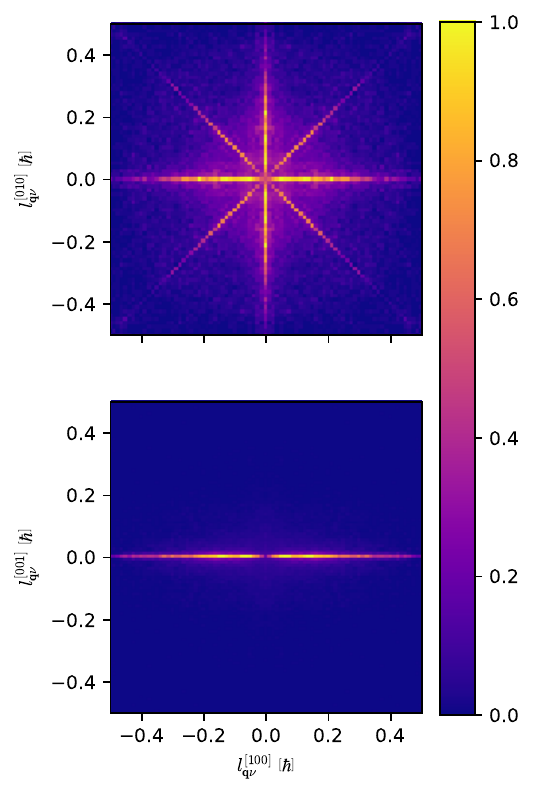}
  \caption{Two-dimensional histograms showing phonon angular momentum distributions in plane perpendicular to $\bm P$ (top) and in plane containing $\bm P$ (bottom) for tetragonal phase of BaTiO$_3$. Data is weighted according to the magnitude of phonon angular momentum. Blue colors indicate phonon angular momenta with few representative phonons, while red and yellow regions indicate regions with many phonons.} 
  \label{fig:dist_planes_tet}
\end{figure}

Now we discuss the distribution of ${\bm l}_{{\bm q} \nu}$ over the phonon wavevectors $\bm{q}$ and branches $\nu$ in the first Brillouin zone.  Figure~\ref{fig:dist_planes_tet} contains two-dimensional histograms showing the fraction of phonons with an angular momentum vector pointing in the directions perpendicular to the polarization (top: $x$-$y$ plane) and the plane containing the polarization direction (bottom: $x$-$z$ plane). From Fig.~\ref{fig:dist_planes_tet}, once again, we see that the phonon angular momentum has a $z$-component that is negligible compared to $x$ and $y$, consistent with our earlier finding.

While the phonon angular momentum is dominantly within the $x$-$y$ plane, we find that there are additional anisotropies within the plane itself.  As can be seen from the top panel of Fig.~\ref{fig:dist_planes_tet} the phonon angular momentum tends to point along specific crystallographic directions. To analyze the anisotropy in the plane perpendicular to $\bm P$ in more detail, we divide phonons into three groups: those with angular momentum pointing within $\pm 5^{\circ}$ of either $[100]$ or $[010]$, those with angular momentum pointing dominantly along $[110]$ or $[\bar{1}10]$ directions, and the remaining phonons. We find that 30\% of the total phonon angular momentum is in the first group, 17\% in the second, while the remaining angular momentum is in the third group. Therefore, nearly half of the phonon angular momentum is located within $\pm 5^{\circ}$ of $[100]$, $[110]$ and symmetry-related directions.

\subsection{Contribution of individual atom types}
\label{sec:individual_atoms}

Next, we analyze the contribution of each atom type to the phonon angular momentum. We focus on angular momentum in the $x$-$y$ plane, perpendicular to $\bm P$, as the angular momentum along $z$ is small.  Given the contribution of the atom $i$ to the angular momentum of the phonon at the wave vector $\bm q$ and branch $\nu$, $\bm{l}^{i}_{{\mathbf{q}}\nu}$, defined in Eq.~\ref{eq:l}, we compute the following average for each atom $i$,
\begin{align}
\label{eq:L_avg_ion}
\big \langle l^{\rm tet}_{\perp} \big \rangle_i
=
\frac{1}{N_q N_{\nu}} \sum_{\bm q} \sum_{\nu} \sqrt{
\left(
 {\bm l}_{{\mathbf{q}}\nu}^i  \cdot {\bm{\hat{n}}}_{100}
\right)^2
+
\left(
{\bm l}_{{\mathbf{q}}\nu}^i \cdot {\bm{\hat{n}}}_{010}
\right)^2
}.
\end{align}
We find that the average contributions of Ba and Ti to the angular momentum are $0.03~\hbar$ and $0.04~\hbar$.  Each of the planar oxygen atoms contributes $0.07~\hbar$ to the average angular momentum.  Apical oxygen has a somewhat smaller contribution $0.05~\hbar$. Therefore, we conclude that about 54\% of the total phonon angular momentum comes from planar oxygen atoms. (We note that the sum of individual atom contributions (0.26~$\hbar$) is larger than total $\langle l_{\perp}^{\rm tet} \rangle$ (0.15~$\hbar$) as contributions from different atoms partially cancel each other.) To check whether averages are different if planar O rotates along the Ti--O bond or perpendicular to it, we separately computed angular momentum averages along $x$ and $y$ directions,
\begin{align}
\big \langle l^{\rm tet}_{x} \big \rangle_i
&=
\frac{1}{N_q N_{\nu}} \sum_{\bm q} \sum_{\nu} \left|
 {\bm l}_{{\mathbf{q}}\nu}^i  \cdot {\bm{\hat{n}}}_{100}
\right|,
\\
\big \langle l^{\rm tet}_{y} \big \rangle_i
&=
\frac{1}{N_q N_{\nu}} \sum_{\bm q} \sum_{\nu} \left|
 {\bm l}_{{\mathbf{q}}\nu}^i  \cdot {\bm{\hat{n}}}_{010}
\right|,
\end{align}
We find that the contribution of the planar oxygen is only slightly larger in the direction along the Ti--O bond ($0.05~\hbar$), than perpendicular to the bond ($0.04~\hbar$).

We also wish to understand the importance of ionic masses on the phonon angular momentum in BaTiO$_3$. Therefore, we decided to change the individual ionic masses used in our calculation by hand.  We keep the force-constant matrices unchanged. We find that the largest increase of the phonon angular momentum ($+30\%$) occurs when we set all masses equal, while the largest decrease of the phonon angular momentum ($-12\%$) corresponds to setting the Ba mass at a value much smaller than that of Ti and O. This is what one would expect based on the argument from the perturbation theory.  For example, if all ionic masses are similar, then one would expect that the atomic vibrations of all ions would have a similar frequency, there would be more hybridization between the phonons, and thus there would be a larger overall angular momentum of the phonon.

\subsection{Relevant displacements of atoms}
\label{sec:relevant_displacmenets}

We have shown in an earlier subsection that planar oxygens are the main contributors to the phonon angular momentum in tetragonal BaTiO$_3$.  Now we focus on determining which atomic displacements $\xi_{\bm q \nu}^{i \alpha}$ are primarily responsible for the angular momentum of the phonon.

We start by defining,
\begin{align}
 c_{{\bm q} \nu} &= 
 \sum_{i\alpha}
 C_{i}^{\alpha} \left \vert {\xi_{\bm q \nu}^{i \alpha}} \right \vert^A ,
 \label{eq:defc}
  \\
d_{{\bm q} \nu}
&=
 \left[ \sum_\alpha \left( l_{\bm q \nu}^\alpha\right)^2 \right]^B.
  \label{eq:defd}
\end{align}
Here $A$, $B$, and $C_{i}^{\alpha}$ are the fitting parameters that we will discuss later. Quantity $c_{{\bm q} \nu}$ is a descriptor of a single phonon mode.  This descriptor depends only on the absolute value of the phonon eigenvector component $\xi_{\bm q \nu}^{i \alpha}$. Therefore, $c_{{\bm q} \nu}$ depends only on the magnitude of the atomic displacements, not on the relative phase between the atomic displacements.  The second quantity, $d_{{\bm q} \nu}$ is simply the square of the norm of the phonon angular momentum vector raised to the $B$-th power.  To establish the relationship, if any, between the two descriptors $c_{{\bm q} \nu}$ and $d_{{\bm q} \nu}$, we seek to find parameters $A$, $B$, and $C_i^{\alpha}$ that minimize the difference between $c_{{\bm q} \nu}$ and $d_{{\bm q} \nu}$.  In other words, we wish to solve the following problem,
\begin{align}
\min_{A, B, C_i^{\alpha}}
\sum_{\bm{q} \nu}
\left(
c_{{\bm q} \nu} 
-
d_{{\bm q} \nu}
  \right)^2.
\label{eq:to_min}
\end{align}
This approach is very similar to the least squares fitting method.  Clearly, if we find $A$, $B$, and $C_i^{\alpha}$ for which there is a good correlation between $c_{{\bm q} \nu}$ and $d_{{\bm q} \nu}$ then a large value of coefficient $C_i^{\alpha}$ can be interpreted as follows: phonons that tend to have a large (in magnitude) displacement of the $i$-th atom in the direction $\alpha$ also tend to have a large angular momentum of the phonon.  Similarly, a small value of $C_i^{\alpha}$ means that displacement of $i$-th atom in direction $\alpha$ does not correlate with the magnitude of the phonon angular momentum.

\begin{figure}[t] 
  \includegraphics{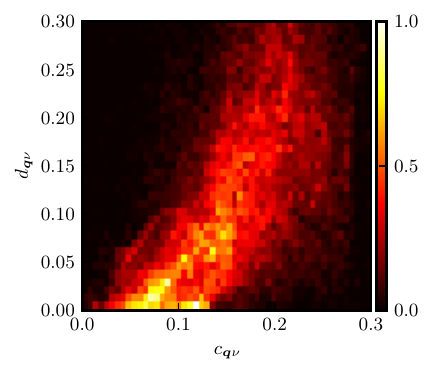}
  \caption{Histogram showing correlation between phonon angular momentum magnitude (vertical axis) and character of atomic motion (horizontal axis) for tetragonal BaTiO$_3$. Quantities on the axes ($c_{{\bm q} \nu}$ and $d_{{\bm q} \nu}$) are explained in the main text. Yellow and red colors represent large density of phonon states while dark red and black bins represent low density.  Data here is shown for optimal values of parameters $A$, $B$, and $C_i^{\alpha}$ used in Eq.~\ref{eq:to_min}.}
  \label{fig:coeffs_t}
\end{figure}

Numerically minimizing Eq.~\ref{eq:to_min} gives optimal values of $A$ and $B$ close to $1/2$.  Optimal values of $C_{i}^{\alpha}$ are given in Table~\ref{tab:c_values}. Figure~\ref{fig:coeffs_t} shows a two-dimensional histogram  indicating a strong correlation between quantities $c_{{\bm q} \nu}$ (horizontal axis) and $d_{{\bm q} \nu}$ (vertical axis). This two-dimensional histogram is constructed by binning individual phonons over all q-points in the dense $30^3$ q-mesh and all fifteen phonon branches $\nu$.   We plot the histogram in Fig.~\ref{fig:coeffs_t} with an optimal choice of $A$, $B$, and $C_{i}^{\alpha}$ we obtained by numerically minimizing Eq.~\ref{eq:to_min}.  Before analyzing numerical values of optimal components $C_i^{\alpha}$, first we briefly discuss  their norm over Cartesian directions. We find that the norm is nearly the same for Ba, Ti, and apical oxygen (0.07), while it is significantly larger (0.16) for each of the two planar oxygens.  This is consistent with the analysis in the previous subsection showing that most of the contribution to the phonon angular momentum comes from planar oxygens.

\begin{table}[!t]
\caption{\label{tab:c_values} 
Optimal values of coefficients $C_i^{\alpha}$ obtained by solving Eq.~\ref{eq:to_min}.  Large positive value of $C_i^{\alpha}$ indicate that large displacement of $i$-th phonon in direction $\alpha$ correlates with large phonon angular momentum. Polarization $\bm P$ points along the $z$-axis.  Data for the planar oxygen in the table corresponds to the planar oxygen for with the Ti--O bond pointing along the $y$-axis.}
\begin{ruledtabular}

\begin{tabular}{
l
d{-2}d{-2}d{-2}
d{-2}d{-2}d{-2}
d{-2}d{-2}d{-2}
d{-2}d{-2}d{-2}
}
& \multicolumn{1}{c}{Ba} & \multicolumn{1}{c}{Ti} & \multicolumn{1}{c}{O-planar} & \multicolumn{1}{c}{O-apical} \\
\hline
$C_i^{x}$ &     0.04 &   -0.01  &   0.05 &   -0.04  \\
$C_i^{y}$ &     0.04 &   -0.01  &   0.01 &   -0.04  \\
$C_i^{z}$ &     0.05 &    0.07  &   0.15 &   -0.02
\end{tabular}

\end{ruledtabular}
\end{table}

Now, we focus on individual coefficients $C_{i}^{\alpha}$  given in Table~\ref{tab:c_values}.  For planar oxygens, we find that the coefficient value is 0.15 for displacement along the $z$ axis, that is, parallel to $\bm P$, while the values for displacements in the $x$-$y$ plane are significantly smaller, 0.05 and 0.01. The smaller value (0.01) corresponds to the displacement of planar oxygen along the Ti--O bond, while 0.05 is for the displacement perpendicular to the bond.

For the Ti atom, we find that the coefficient $C_{i}^{\alpha}$ for displacement along the $\bm P$ direction is $0.07$, while the in-plane displacements of the titanium atom are effectively uncorrelated, since the corresponding coefficients are $-0.01$. Coefficient $C_{i}^{\alpha}$ for Ba atom both parallel to $\bm P$ and in-plane is 0.04. The apical oxygen atom has a relatively weak but negative coefficient of $-0.04$  for displacements in the $x$ and $y$ directions.

Therefore, we conclude that phonons that predominantly involve displacements of planar oxygen and titanium atoms parallel to $\bm P$ are the ones that are the most correlated with a large angular momentum.  Clearly, modes that have phonon angular momentum must also involve motion in some direction perpendicular to $\bm P$, as otherwise Eq.~\ref{eq:l} would give zero angular momentum.  

\subsection{Example: modes near inversion symmetric points}\label{subsec:example_modes}

After analyzing the distribution of the phonon angular momentum over the entire Brillouin zone, we now focus on one representative region of the Brillouin zone, near the $X$ point.  In the cubic phase of high symmetry, there are three $X$ points which all map onto themselves under inversion.  At each $X$ point, there are five double-degenerate phonon branches, whereas the remaining phonons are nondegenerate.  In the tetragonal phase, with the polar axis chosen along the $z$-axis, the three equivalent face centers are reduced to two. At points $[1/2,0,0]$ and $[0,1/2,0]$ each of the five previously doubly degenerate branches is split. We choose to study as an example one of the branches that dominantly came from previously doubly degenerate modes.\footnote{For purposes of this example calculation, we used here structure of BaTiO$_3$ that is interpolated between cubic and tetragonal phase.  More precisely, we set $\lambda^{\rm tet}$ parameter (defined in Sec.~\ref{sec:saturation}) to $0.2$.}  The frequency of this particular mode is 9.4~THz and it mainly involves motions of oxygen atoms.  We parameterize the linear momentum of phonons in the vicinity of the high-symmetry point as, $$\bm{q}=[\epsilon_x+1/2,~\epsilon_y,~\epsilon_z]$$ for a small value of $\epsilon_x$, $\epsilon_y$, and $\epsilon_z$. The calculated corresponding phonon angular momentum is $$(l_x,l_y,l_z)\sim (5\epsilon_y,\epsilon_x,0)$$ within the first order in $\epsilon$'s. The linear dependence of $l$ on $\epsilon$ is a consequence of the fact that a given phonon branch at $\bm q$ and $-\bm q$ has opposite phonon angular momentum.  Therefore, at lowest order in the Taylor expansion we expect that $l$ scales with a first power of $\epsilon$. 

\begin{figure}[t]
  \includegraphics{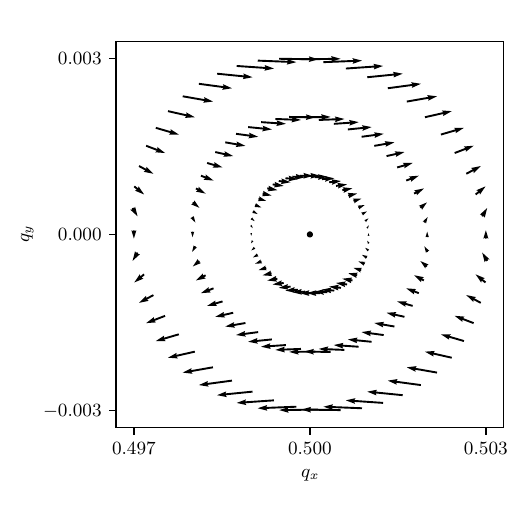}
  \caption{Phonon angular momentum vectors in the vicinity of  $q=\left[ \frac{1}{2}~0~0 \right]$ for one of the optical branches in tetragonal BaTiO$_3$ (see text for more detail). The arrows are proportional to the phonon angular momentum with the largest arrow corresponding to $0.041~\hbar$.}
  \label{fig:arrows}
\end{figure}

Furthermore, we find that  the $l_x$ Cartesian component is proportional to $\epsilon_y$ while $l_y$ component is proportional to $\epsilon_x$.  Therefore, if one moves away from the high-symmetry point, the phonon angular momentum winds around it once. The phonon angular momentum given in the above equation is indicated by the black arrows in Fig.~\ref{fig:arrows}.

\section{Saturation of phonon angular momentum}
\label{sec:saturation}

In the previous section, we focused on the dependence of the phonon angular momentum on the direction of polar atomic displacements, and thus polarization $\bm{P}$.  Now we analyze the dependence of the phonon angular momentum on the magnitude of the polar atomic displacements. To check the dependence of phonon angular momentum on the polar displacement magnitude, we linearly transform the crystal structure from the cubic phase to the tetragonal phase. We parameterize structures between cubic and tetragonal with the parameter $\lambda^{\rm tet}$. By definition, when $\lambda^{\rm tet}=0$ atom positions (free parameters in Wyckoff orbits) and lattice parameters correspond to the cubic phase.  Similarly, $\lambda^{\rm tet}=1$ corresponds to the polar tetragonal phase.  When $0<\lambda^{\rm tet}<1$ the structural parameters are linearly interpolated between the cubic and polar tetragonal phases.

For small atomic displacements, and thus small $\lambda^{\rm tet}$, the phonon angular momentum in BaTiO$_3$ is by symmetry linearly proportional to polar atomic displacement parameterized by $\lambda^{\rm tet}$. This is easy to see, as parameter $\lambda^{\rm tet}$, polarization ${\bm P}$, and phonon angular momentum ${\bm l}_{{\bm q} \nu}$ all change sign under inversion symmetry present in the bulk.  Therefore, in the lowest order of Taylor expansion ${\bm l}_{{\bm q} \nu}$ is proportional to the first power of $\lambda^{\rm tet}$.  Since ${\bm l}_{{\bm q} \nu} \sim \lambda^{\rm tet},$ one might hope that in some hypothetical material with even larger polar displacement than BaTiO$_3$ one might find even larger phonon angular momentum.  Nevertheless, our analysis shows that this scenario is unlikely to happen, at least not in ABO$_3$ perovskites.

\begin{figure*}[htp]
  \includegraphics{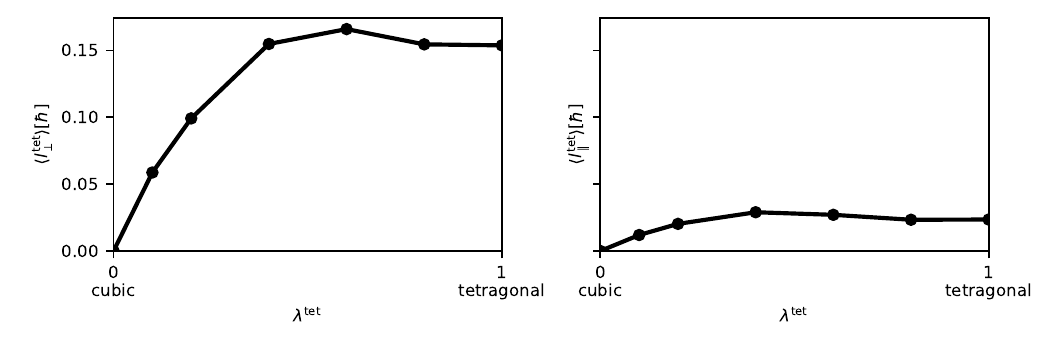}
  \caption{Phonon angular momentum average in tetragonal BaTiO$_3$ reaches 15\% of its maximum possible value. Horizontal axis ($\lambda^{\rm tet}$) parameterizes the atomic displacements, along with changes in the lattice constants, as BaTiO$_3$ transitions from cubic ($\lambda^{\rm tet}$= 0) to polar tetragonal ($\lambda^{\rm tet}$= 1). Vertical axis shows  calculated average angular momentum per phonon, in a tetragonal phase of BaTiO$_3$, as a function of $\lambda^{\rm tet}$ in directions perpendicular (left) to $\bm P$ and parallel (right) to $\bm P$.}
  \label{fig:pam_lambda}
\end{figure*}

Figure~\ref{fig:pam_lambda} shows the averaged phonon angular momentum as a function of $\lambda^{\rm tet}$ between 0 and 1. The left panel of Fig.~\ref{fig:pam_lambda} shows the average phonon angular momentum in the plane perpendicular to $\bm P$, while the right panel shows the average phonon angular momentum along $\bm P$. By power-law fitting for small $\lambda^{\rm tet}$ we find that the average angular momentum is linearly proportional to small Ti displacement.

For $\lambda^{\rm tet}$ above 0.5 the angular momentum averages saturate to a constant value and do not change significantly  as $\lambda^{\rm tet}$ is increased from 0.5 to 1.0. 
As we are about to see in section~\ref{sec:anisotropies}, when $\lambda^{\rm tet}$ is around 0.5 some of the Ti--O bonds break and remain broken in the entire range from 0.5 to 1.0.  We speculate that the phonon angular momentum saturation is the result of these Ti--O bonds breaking at $\lambda^{\rm tet} \approx 0.5$.

Therefore, we expect that phonon angular momentum in a ferroelectric such as BaTiO$_3$ can't be increased further by simply increasing the polar displacement.  

We note that the ratio between the angular momentum in the plane and parallel to $\bm P$ is around 5--6, regardless of the value of $\lambda^{\rm tet}$.

\section{Origin of anisotropies}\label{sec:anisotropies}

In Sec.~\ref{sec:tet} we have shown that the phonon angular momentum in tetragonal BaTiO$_3$ is asymmetric in two ways. First, the angular momentum is about 6 times greater in the plane perpendicular to polarization $\bm P$ than along $\bm P$.  Second, the angular momentum in the plane perpendicular to $\bm P$ is significantly stronger along certain crystallographic directions. In this section, we study the origin of these anisotropies by studying the anisotropy in the calculated interatomic force constants.

The interatomic force constants $F^{\alpha\beta}_{ij}(\bm R)$ are composed of both short-range interactions ($F_{\rm SR}$) and long-range dipole-dipole interactions ($F_{\rm DD}$).\cite{Gonze1994} However, it is unclear in our case whether short-range or long-range interactions are more relevant for the calculated phonon angular momentum.  For this reason, we performed a hybrid calculation, in which we set the dipole-dipole interaction strength to zero and recalculated the angular momentum of all phonons. Technically, we did this by setting the diagonal components of the electronic part of the electron permittivity $\epsilon_\infty$ to infinity instead of using the calculated value.  With this hybrid approach, we find that $\big \langle l_{\parallel}^{\rm tet} \big \rangle$ slightly increases from $0.02~\hbar$ to $0.04~\hbar$, while $\big \langle l_{\perp}^{\rm tet} \big \rangle$ decreases from $0.15~\hbar$ to $0.11~\hbar$. Therefore, the angular momentum is still anisotropic, and we are justified in focusing on the short-range part of the interatomic force constant.  Furthermore, to simplify the analysis, we don't consider here the entire $3 \times 3$ matrix $F^{\alpha\beta}_{ij}(\bm R)$ but rather the magnitude $k_{i}^{j{\bm R}}$ summed over Cartesian directions, $\sum_{\alpha\beta}|F^{\alpha\beta}_{ij}(\bm R)|$.  We use $k_{i}^{j{\bm R}}$ to measure the strength of interatomic forces between the atom $i$ in the home cell $\bm{R}=0$ and the atom $j$ translated by the lattice vector ${\bm R}$.

\begin{figure}[htp]
  \includegraphics{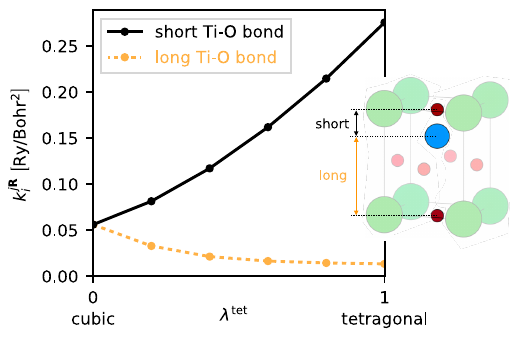}
  \caption{Magnitude of interatomic forces $k_{i}^{j{\bm R}}$ between titanium and the nearest neighboring apical oxygen atoms as a function of $\lambda^{\rm tet}$ (defined in Sec.~\ref{sec:saturation}). Solid black line corresponds to the Ti--O bond which shortens with $\lambda^{\rm tet}$. Orange color corresponds to the elongated Ti--O bond. These bonds are also indicated in the inset. For $\lambda^{\rm tet}$ above roughly $0.5$, the elongated bond is effectively broken as calculated $k_{i}^{j{\bm R}}$  tends to zero. Indices $i$, $j$, and lattice vector $\bm R$ are selected so that the value of $k_{i}^{j{\bm R}}$ used in the plot corresponds to the pairs of nearest neighboring Ti and apical O atoms.}
  \label{fig:TiO_ifcs}
\end{figure}

First, we study the interatomic force strengths in the cubic phase and then compare them to those in the polar tetragonal phase.  In the cubic phase, the strongest interaction strength $k_{i}^{j{\bm R}}$ is $0.06~\rm{Ry/Bohr}^2$, corresponding to the nearest neighboring Ti and O atoms.  This observation is consistent with the fact that there is a strong covalent-like bond between the nearest neighboring Ti and O atoms. In the tetragonal phase, with Ti displaced along $\bm P$, the strength of interaction between Ti and planar oxygens increases from $0.06~\rm{Ry/Bohr}^2$ to $0.09~\rm{Ry/Bohr}^2$. As shown in Fig.~\ref{fig:TiO_ifcs}, the change in the interaction with apical oxygen is even more drastic, since the displacement of the Ti atom significantly changes the length of the bond to apical oxygen. The strength of the interaction between Ti and the apical O with a short bond increases to $0.28~\rm{Ry/Bohr}^2$ (solid line in Fig.~\ref{fig:TiO_ifcs}) while the interaction along the elongated bond (in the home cell $\bm{R}=0$) is reduced to only $0.005~\rm{Ry/Bohr}^2$ (dashed line in Fig.~\ref{fig:TiO_ifcs}), as expected for an effectively broken covalent-like bond. Because one of the apical oxygens effectively does not interact with the Ti atom, we conclude that the covalently bonded three-dimensional network of Ti and O atoms in the cubic phase has been essentially reduced to a quasi two-dimensional network of Ti and O atoms in the tetragonal phase. Next, we analyze the character of the interatomic forces within the two-dimensional plane of atoms.  As discussed in Sec.~\ref{sec:relevant_displacmenets} the phonon angular momentum dominantly comes from a motion of Ti and planar O atoms.  Therefore, we don't include in the discussion Ba or apical O atoms.

\begin{figure}[htp] 
  \includegraphics{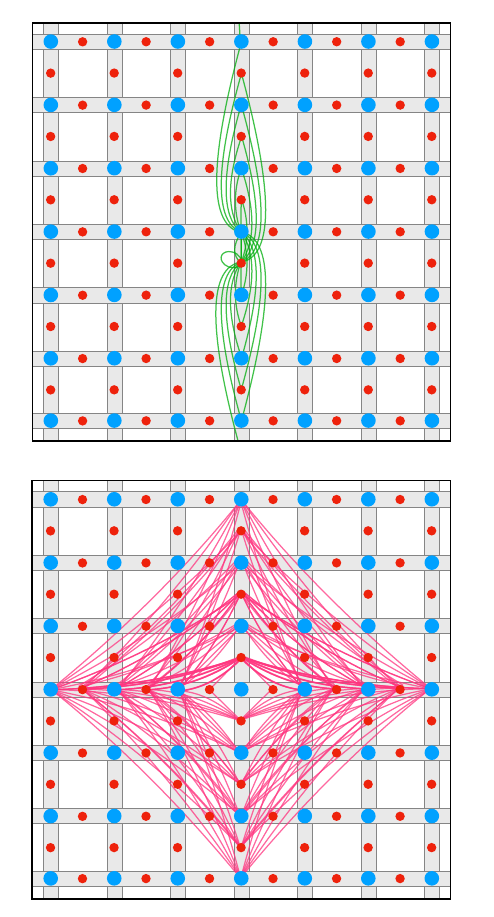}
  \caption{Planar oxygen atoms are shown with solid red circles while titanium atoms are shown with solid blue circles.  We find that the atomic interactions in BaTiO$_3$ are strong along the $\ldots$--Ti--O--Ti--O--$\ldots$ chains.  These chains are indicated with gray color. To characterize the strength of the bonds within the chain we summed the force constant matrix elements ($k_{i}^{j{\bm R}}$) for all pair of atoms in the same chain (pair are indicated with green lines in the top panel).  The resulting sum is $0.92~\rm{Ry/Bohr}^2$ per one atom. Interactions between two chains are indicated with pink lines (bottom panel).  These interactions are about three times smaller, as sum of $k_{i}^{j{\bm R}}$ adds up to $0.31~\rm{Ry/Bohr}^2$ per one atom.}
  \label{fig:2by2}
\end{figure}

Now we further decompose the quasi two-dimensional network of Ti and planar O atoms.  We consider the planar network of Ti and O atoms as a series of separate subsystems, each consisting of infinite one-dimensional chains of Ti and planar O atoms.  These chains are indicated in gray in Fig.~\ref{fig:2by2}. With such a decomposition, we can now quantify the interatomic forces within a single chain and between chains. To quantify interactions along a single $\ldots$--Ti--O--Ti--O--$\ldots$ chain of atoms, we sum $k_{i}^{j{\bm R}}$ over all $i$ and $j{\bm R}$ corresponding to the same chain.  These interactions are shown by green lines from atom $i$ to atom $j{\bm R}$ in the top panel of Fig.~\ref{fig:2by2}. The resulting sum is equal to $0.92~\rm{Ry/Bohr}^2$. Next, we consider a perpendicular pair of chains and sum $k_{i{\bm R}^\prime}^{j{\bm R}}$ over all pairs where $i {\bm R}^\prime$ and $j{\bm R}$ correspond to atoms in different chains. These interactions are shown as pink lines connecting atoms $i {\bm R}^\prime$ and $j{\bm R}$ as in the bottom panel of Fig.~\ref{fig:2by2}.  This results in a value about three times smaller, $0.31~\rm{Ry/Bohr}^2$.\footnote{The remaining interatomic forces, not included in the sums above, are the onsite terms where both $i$ and $j{\bm R}$ correspond to the atoms in the home cell $\bm{R}=0$. These on-site terms for both Ti and O sum to $0.54~\rm{Ry/Bohr}^2$.}  Therefore, we conclude that the interatomic forces are effectively strong within the chain and weak between the chains. In other words, the interatomic forces in BaTiO$_3$ effectively have low dimensionality.  This crystal can be seen as consisting of strongly bonded one-dimensional $\ldots$--Ti--O--Ti--O--$\ldots$ chains that are connected to each other and form a quasi two-dimensional network. 

The quasi-one-dimensionality of the force constant matrix is consistent with our earlier finding that the angular momentum of the phonon is dominantly pointing along the specific crystallographic directions perpendicular to $\bm P$.  In the following section, we introduce an analytical model to give a simple physical picture of this finding.

\subsection{Analytical model}
\label{sec:model}

Now, we analytically study the phonon angular momentum for a model of a chain of repeating O and Ti atoms. We represent interatomic interactions with springs between the nearest neighboring Ti and O atoms. Each spring is characterized by two spring constants: one for stretching ($K_{\rm r}$) and one for bending ($K_\theta$). The potential energy summed over the nearest-neighbor interactions $\langle ij \rangle$ is~\cite{kaxiras} 
\begin{equation}\label{eq:potential}
V=\frac{1}{2} \sum_{\langle ij \rangle} \left[ \left( K_{\rm r} - K_\theta \right)\left[ {\bf s}_{ij} \cdot {\bf \hat{r} }_{ij} \right]^2 + K_\theta \left| {\bf s}_{ij}\right|^2 \right].
\end{equation}
Here, ${\bf \hat{r} }_{ij}$ is the unit vector connecting the atoms $i$ and $j$. The displacement of the atom $i$ is ${\bf s}_i$, while ${\bf s}_{ij}$ is defined as ${\bf s}_j - {\bf s}_i$.
\begin{figure}[!t]
  \includegraphics[width=3.4in]{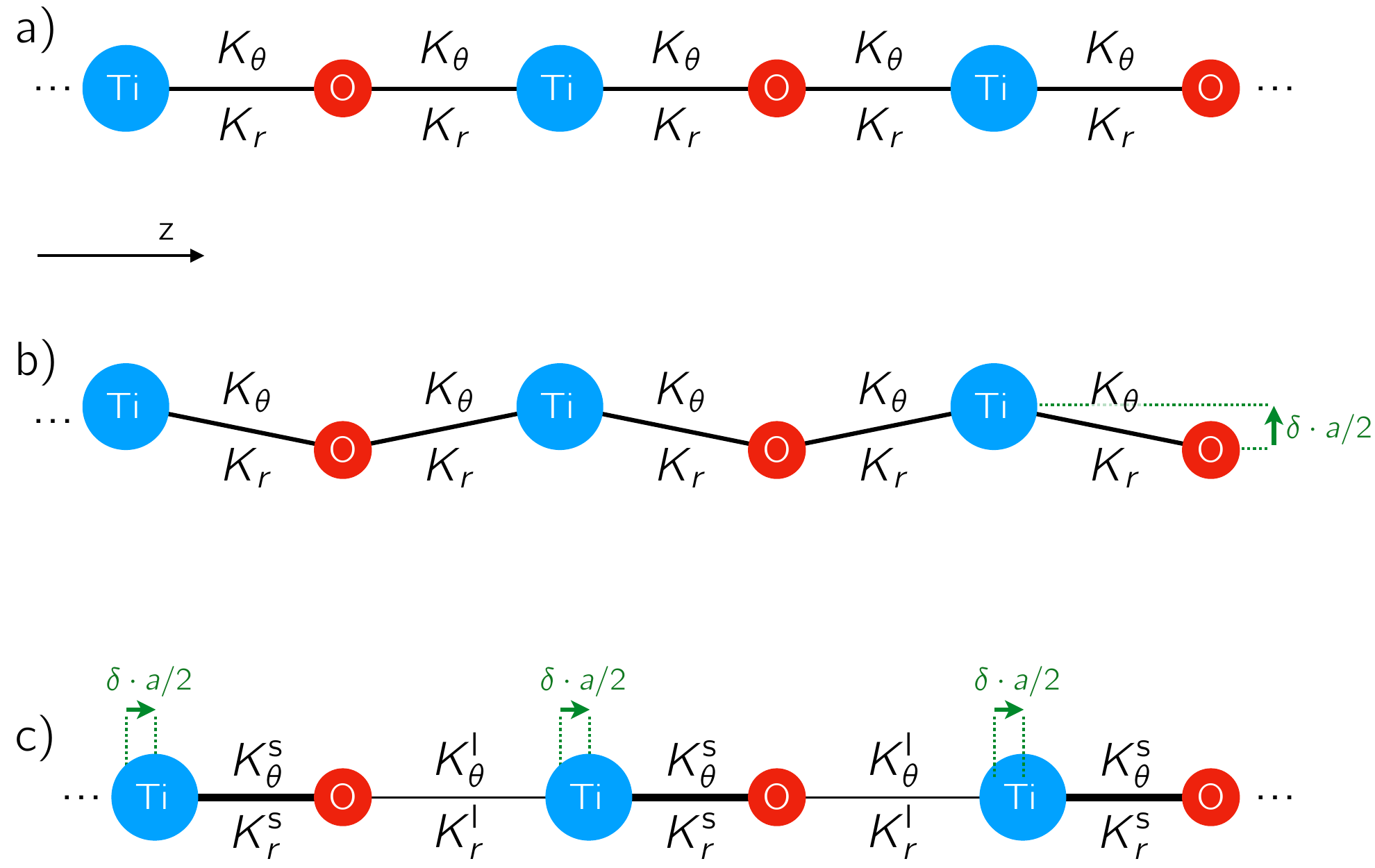}
  \caption{Schematics of various one-dimensional models we studied. Blue circles are titanium atoms and red circles are oxygen atoms. Atoms are connected by springs shown as black lines.  Each spring includes potential energy term for both stretching ($K_{\rm r}$) and bending ($K_\theta$), as described in the main text. Displacements of atoms in panels b and c are indicated with green arrow. In panels a and c phonons can be chosen so that $l^{i \alpha}_{{\bm q} \nu}=0$. Displacement of atoms in direction perpendicular to the chain (case b) generates phonon angular momentum $l^{ix}_{{\bm q} \nu}$ on titanium and oxygen atom which points perpendicular to both chain direction and the atom displacement (in and out of page).  Nevertheless, $l^x_{{\bm q} \nu} = \sum_i l^{ix}_{{\bm q} \nu} = 0$ even in the case of panel b, as the contribution of titanium atom cancels that of the oxygen atom.  This cancellation does not occur in the two-dimensional extension of the model discussed in Sec.~\ref{sec:model2d} or when further neighbors are included in the model.\cite{supplement}}
  \label{fig:model}
\end{figure}
\begin{figure}[htp]
  \includegraphics[width=3.4in]{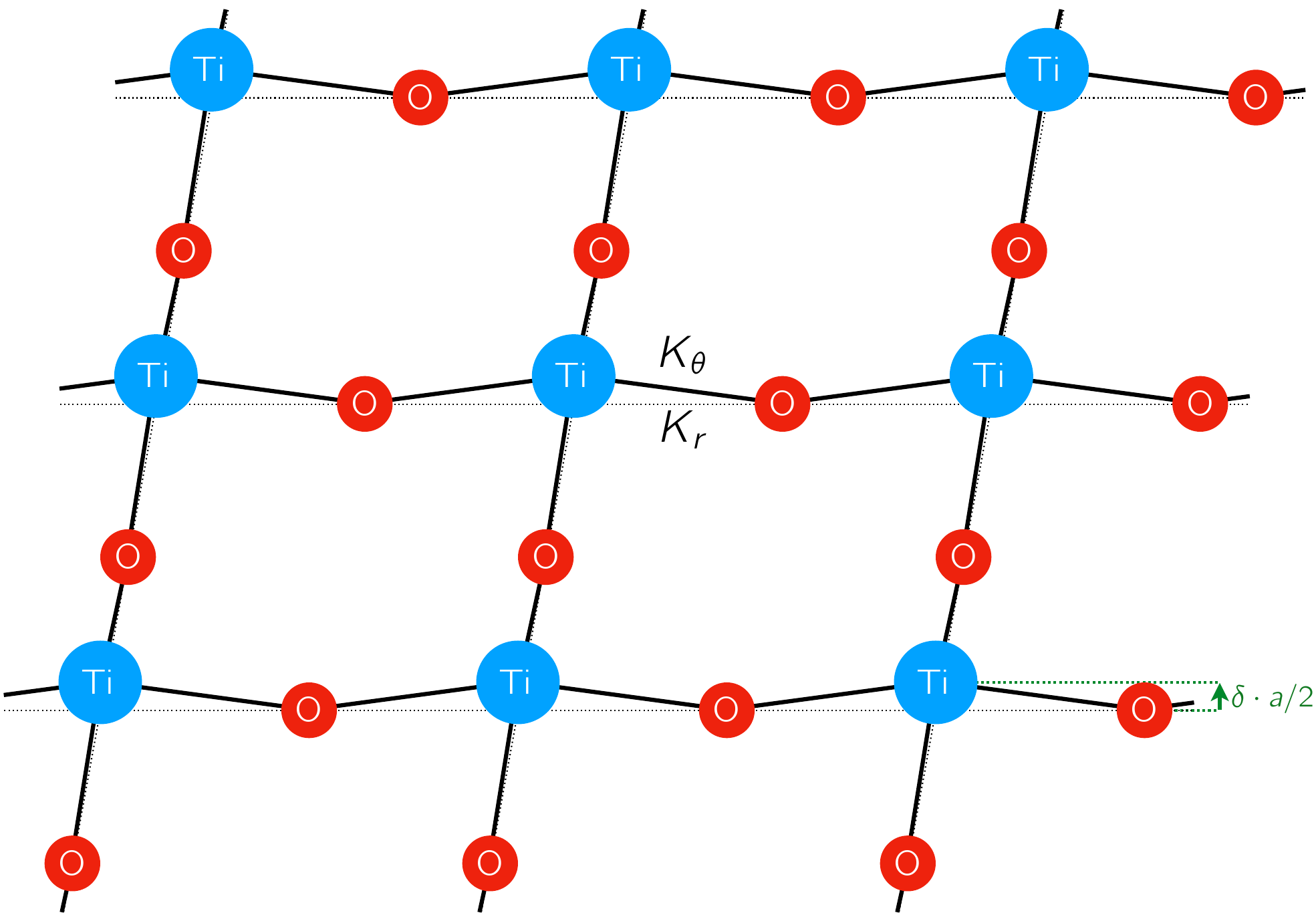}
  \caption{Two-dimensional model we studied.  Conventions in the figure are the same as in Fig.~\ref{fig:model}.  Titanium and oxygen atoms form a two-dimensional plane.  Titanium atoms are displaced by $\delta a /2$ along the third dimension, perpendicular to the two-dimensional plane of atoms.  Generically, phonons in this model have non-zero phonon angular momentum.  See text for more details.}
  \label{fig:model2d}
\end{figure}
We use the potential from Eq.~\ref{eq:potential} to derive the dynamical matrices following a standard approach,  
\begin{equation}\label{eq:dynmat_general}
D_{ij}^{\alpha\beta}(q)=\sum_n {\rm e}^{-i q R_n} \frac{1}{\sqrt{M_{i} M_{j}}}\frac{\partial^2 V}{\partial s_{ n i\alpha}\partial s_{0 j\beta}}.
\end{equation}
Here, $R_n$ is the location of the $n$-th unit cell and $q$ is the wavevector.  We analytically computed the dynamical matrices for the models shown in Figs.~\ref{fig:model} and \ref{fig:model2d}.  As an additional test, we also constructed these models using a general-purpose computer package given in the supplement.\cite{supplement}

In our one-dimensional chain with two atoms per unit cell, the dynamical matrix is a $6 \times 6$ matrix, as we allow each atom to move along all three Cartesian directions. We assign the indices of this matrix so that the first and second $2 \times 2 $ sub-blocks of the dynamical matrix correspond to the atomic displacements perpendicular to the chain (directions $x$ and $y$), while the third sub-block corresponds to movements along the chain (direction $z$),
\begin{align}
{\bm D } =
\begin{bmatrix}
{\bm D}^{xx \phantom{\dagger}} & {\bm D}^{xy \phantom{\dagger}} & {\bm D}^{xz} \\
{\bm D}^{xy \dagger} & {\bm D}^{yy \phantom{\dagger}} & {\bm D}^{yz} \\
{\bm D}^{xz \dagger} & {\bm D}^{yz \dagger} & {\bm D}^{zz} \\
\end{bmatrix}.
\end{align}
(Here we are using the fact that $\bm D$ is a Hermitian matrix which implies for the $2 \times 2$ sub-blocks that ${\bm D}^{\alpha \beta} = {\bm D}^{\beta \alpha \dagger}$.)
Each $2 \times 2$ sub-block ${\bm D}^{\alpha \beta}$ is arranged so that the indices correspond to the displacements of the titanium and the oxygen atom as follows,
\begin{align}
{\bm D}^{\alpha \beta}  =
\begin{bmatrix}
D_{{\rm Ti \, Ti}}^{\alpha \beta} & D_{{\rm Ti \, O}}^{\alpha \beta} \\
D_{{\rm O \, Ti}}^{\alpha \beta} & D_{{\rm O \, O}}^{\alpha \beta} \\
\end{bmatrix}.
\end{align}

\subsubsection{Model shown in Fig.~\ref{fig:model}(a)}

We start by analyzing the system shown in Fig.~\ref{fig:model}(a). This system consists of an equidistant chain of Ti and O atoms connected by springs that can stretch ($K_{\rm r}$) and bend ($K_{\theta}$).  Following Eq.~\ref{eq:potential} the potential energy of this system is given by,
\begin{align}
V=\frac{1}{2}\sum_{\langle ij \rangle}\big[K_{\rm r} s_{ijz}^2 + K_{\theta}(s_{ijx}^2+s_{ijy}^2)\big].
\label{eq:model8a}
\end{align}
Performing the sum over the nearest neighbors and computing the dynamical matrix gives,
\begin{equation}
{\bm D} =
\begin{bmatrix}
\pmb{A}_1 & {\bm 0} & {\bm 0}\\
{\bm 0} & \pmb{A}_1 & {\bm 0}\\
{\bm 0} & {\bm 0} & \pmb{A}_2\\
\end{bmatrix}. \label{eq:simple}
\end{equation}
The $2 \times 2$ sub-matrices $\pmb{A}_1$ and $\pmb{A}_2$ are,
\begin{equation}
 \pmb{A}_1 = K_\theta\begin{bmatrix}
\frac{2}{M_{\rm Ti}} & \frac{-1-e^{-iqa}}{\sqrt{M_{\rm Ti}M_{\rm O}}}\\
 \frac{-1-e^{iqa}}{\sqrt{M_{\rm Ti}M_{\rm O}}} & \frac{2}{M_{\rm O}}
\end{bmatrix}
\end{equation}
\begin{equation}
\pmb{A}_2 = \frac{K_{\rm r}}{K_\theta}\pmb{A}_1
\end{equation}
$M_{\rm Ti}$ and $M_{\rm O}$ are the atomic masses of titanium and oxygen, $a$ is the lattice constant, and $q$ is the phonon wave vector. As usual, we compute the phonon eigenvectors $\xi_{i\alpha}$ via $$\sum_{j, \beta}D_{ij}^{\alpha \beta}(\bm q) \xi_{j\beta}=\omega^2\xi_{i\alpha}.$$

Dynamical matrix from Eq.~\ref{eq:simple} is block-diagonal in the Cartesian indices.  In other words,
\begin{align}
{\bm D}^{xy} =  {\bm D}^{xz} =  {\bm D}^{yz} = {\bm 0}.
\end{align}
Therefore, the eigenvectors of Eq.~\ref{eq:simple} can be chosen to correspond to atomic motions along only one of the Cartesian axes (since a block-diagonal matrix can effectively be diagonalized one block at a time).  Therefore, clearly, the corresponding atomic motions are collinear and the phonon angular momentum is zero.  \footnote{It is clear from Eq.~\ref{eq:summed_ion_ang} that angular momentum $l_z=0$ unless both $x$ and $y$ components of phonon eigenvector $\xi$ are non-zero.} This finding is consistent with the fact that our model is symmetric under inversion. \footnote{With Fig.~\ref{fig:model}(a) in mind, one can select the center of any atom or spring as the origin, apply the inversion operator ($z\rightarrow -z$), and find the system unchanged.}

\subsubsection{Model shown in Fig.~\ref{fig:model}(b)}

Next, we break the inversion symmetry in the model by slightly displacing Ti in the direction perpendicular to the chain, as in Fig.~\ref{fig:model}(b). The magnitude of the Ti displacement is $\delta \cdot a/2$, where $a$ is the lattice constant. There are two directions perpendicular to the chain, and to be precise, we chose to move the atom along the direction $y$ corresponding to the second column (row) of the dynamical matrix. For such a system $\pm {\bf \hat{r} }_{ij} = \frac{\delta}{\sqrt{1+\delta^2}}\hat{y}+\frac{1}{\sqrt{1+\delta^2}}\hat{z}$. If we again assume that nearest neighboring Ti and O atoms are connected by springs with bond stretching ($K_{\rm r}$) and bond bending
($K_\theta$) terms, the potential energy for Fig.~\ref{fig:model}(b) is
\begin{align}
V= & \frac{1}{2} \sum_{\langle ij\rangle} 
\frac{K_{\rm r} - K_{\theta}}{(1+\delta^2)^2} 
\left[ s_{ijz}^2 + \delta^2 s_{ijy}^2 + 2 \delta (1+\delta^2) s_{ijy}s_{ijz} \right]
\notag \\
+ &
\frac{1}{2} \sum_{\langle ij\rangle} K_{\theta} \left( s_{ijx}^2+s_{ijy}^2+s_{ijz}^2 \right),
\end{align}
Therefore, compared to Eq.~\ref{eq:model8a}, the leading order correction to the ion dynamics is linear in $\delta$, and equals
\begin{align}
\delta \sum_{\langle ij\rangle} 
(K_{\rm r} - K_{\theta}) s_{ijy} s_{ijz} + {\cal O} (\delta^2)
\label{eq:correction}
\end{align}
As can be seen from the functional form of this term, this interaction will lead to coupling of the atomic motion in the direction perpendicular to the chain ($y$) and in the direction along the chain ($z$). 

Calculating the dynamical matrix for this model, up to all orders in $\delta$, gives us
\begin{equation}
\begin{bmatrix}
\pmb{A}_1 & {\bm 0}  &{\bm 0} \\
{\bm 0} & \pmb{A}_1^\prime & \pmb{B}^\prime\\
{\bm 0} & \pmb{B}^\prime & \pmb{A}_2^\prime
\end{bmatrix}. \label{eq:perp}
\end{equation}
Here $\pmb{A}_1^\prime$, $\pmb{A}_2^\prime$, and $\pmb{B^\prime}$ are defined as
\begin{align*}
\pmb{A}_1^\prime &=
\left[
1+
\frac{\delta^2}{1+\delta^2}
\left(\frac{K_{\rm r}}{K_\theta}-1\right)
\right]
\pmb{A}_1,
\\~\\
\pmb{A}_2^\prime &=
\frac{1}{1+\delta^2} \left( \frac{K_{\rm r}}{K_\theta} + \delta^2 \right)
\pmb{A}_1,
\\~\\
\pmb{B}^\prime &=
\frac{\delta}{1+\delta^2}\left(K_{\rm r} - K_\theta \right)
\begin{bmatrix}
0 & \frac{1-e^{-iqa}}{\sqrt{M_{\rm Ti}M_{\rm O}}}\\~\\
 \frac{1-e^{iqa}}{\sqrt{M_{\rm Ti}M_{\rm O}}} & 0
\end{bmatrix}  .
\end{align*}
As can be seen from Eq.~\ref{eq:perp} some of the off-diagonal sub-matrices are zero,
\begin{align}
{\bm D}^{xy} = {\bm D}^{xz} = {\bm 0}.
\label{eq:db1}
\end{align}
However, the off-diagonal sub-matrix coupling motion in the $y$ and $z$ directions is nonzero, 
\begin{align}
{\bm D}^{yz}=\pmb{B}^\prime \neq {\bm 0}
\label{eq:db2}
\end{align}
as expected from the functional form of Eq.~\ref{eq:correction}. Therefore, diagonalizing Eq.~\ref{eq:perp}, corresponding to model Fig.~\ref{fig:model}(b), generally results in phonon eigenvectors in which atoms are allowed to move in the entire $y$-$z$ plane. For a low symmetry $q$ (that is not an integer multiple of $\pi/a$) we find that the atomic motions of the Ti and O atoms are elliptical, so that $l^{ix}_{{\bm q} \nu}$ is generally non-zero.  The other two components are zero, $l^{iy}_{{\bm q} \nu}=l^{iz}_{{\bm q} \nu}=0$. This is to be expected from Eqs.~\ref{eq:db1} and \ref{eq:db2} since the only non-zero off-diagonal sub-matrix is ${\bm D}^{yz}$.  In other words, the angular momentum of the phonon is perpendicular both to the chain direction ($z$) and to the direction of the displacement of the atom ($y$).  Nevertheless, even though  $l^{ix}_{{\bm q} \nu}$ is non-zero, the total angular momentum $l^x_{{\bm q} \nu} = \sum_i l^{ix}_{{\bm q} \nu}$ is zero, as the contribution from the two atoms in the unit cell cancels out.  As we will see in Sec.~\ref{sec:model2d}, this cancelation is not present in the extension of this model to two dimensions, or when springs between further neighboring atoms are included in the model.\cite{supplement}

\subsubsection{Model shown in Fig.~\ref{fig:model}(c)}

Next, for completeness, we also studied our model when  Ti is displaced along the chain, as shown in Fig.~\ref{fig:model}(c). The system is once again one-dimensional and ${\bf \hat{r} }_{ij} = \pm \hat{z}$. Since the Ti--O distances are now not the same, we parameterize the stretching of the short Ti--O bond with $K^{\rm s}_{r}$ and the long Ti--O bond with $K^{\rm l}_{r}$. Similar for bending constants $K^{\rm s}_\theta$ and $K^{\rm l}_\theta$. The resulting dynamical matrix is
\begin{equation}
\begin{bmatrix}
\pmb{A}^{\prime\prime}_1 & \bm{0} & \bm{0} \\
\bm{0}  & \pmb{A}^{\prime\prime}_1 & \bm{0} \\
\bm{0}  & \bm{0}  & \pmb{A}^{\prime\prime}_2
\end{bmatrix}, \label{eq:along}
\end{equation}
where the matrices $\pmb{A}^{\prime\prime}_1$ and $\pmb{A}^{\prime\prime}_2$ are defined as
\begin{align*}
\pmb{A}_1^{\prime\prime} &= \begin{bmatrix}
\frac{K^{\rm s}_{\theta}+K^{\rm l}_{\theta}}{M_{\rm Ti}} & \frac{-K^{\rm s}_{\theta}-K^{\rm l}_{\theta} e^{-iqa}}{\sqrt{M_{\rm Ti}M_{\rm O}}}\\~\\
 \frac{-K^{\rm s}_{\theta}-K^{\rm l}_{\theta} e^{iqa}}{\sqrt{M_{\rm Ti}M_{\rm O}}} & \frac{K^{\rm s}_{\theta}+K^{\rm l}_{\theta}}{M_{\rm O}}
\end{bmatrix},
\end{align*}
\begin{align*}
\pmb{A}_2^{\prime\prime} &= \begin{bmatrix}
\frac{K^{\rm s}_{r}+K^{\rm l}_{r}}{M_{\rm Ti}} & \frac{-K^{\rm s}_{r}-K^{\rm l}_{r} e^{-iqa}}{\sqrt{M_{\rm Ti}M_{\rm O}}}\\~\\
 \frac{-K^{\rm s}_{r}-K^{\rm l}_{r} e^{iqa}}{\sqrt{M_{\rm Ti}M_{\rm O}}} & \frac{K^{\rm s}_{r}+K^{\rm l}_{r}}{M_{\rm O}} .
\end{bmatrix}
\end{align*}
Since the dynamical matrix from Eq.~\ref{eq:along} satisfies 
$$
{\bm D}^{xy} =  {\bm D}^{xz} =  {\bm D}^{yz} = {\bm 0}
$$ 
we conclude that $l^{i\alpha}_{{\bm q} \nu}=0$ for the model in Fig.~\ref{fig:model}(c), as all phonon modes can again be chosen to consist of collinear atomic motion.

While the models presented so far consist of effective interatomic springs only between the first nearest neighbors, most features of the model are unchaged even when further neighbors are included in the model. We provide more details on these models in the supplement.\cite{supplement}

\subsubsection{Two-dimensional model shown in Fig.~\ref{fig:model2d}}
\label{sec:model2d}

Now we study the generalization of our model to two dimensions.  This model is shown in Fig.~\ref{fig:model2d} and consists of a two-dimensional plane of titanium and oxygen atoms.  There are now three atoms in the primitive unit cell (one titanium atom and two oxygen atoms). The titanium atom is displaced along the third dimension, perpendicular to the two-dimensional plane of atoms.  The numerical implementation of the dynamical matrix of this model is given in the supplement.\cite{supplement} Following the same procedure as in the previous models, we diagonalize the dynamical matrix and compute the angular momentum of the phonon.  As expected, we find that the phonon angular momentum points in the two-dimensional plane of atoms. The total phonon angular momentum $l^{\alpha}_{{\bm q} \nu}$ is now non-zero, as contributions from three atoms in the unit cell $ l^{i \alpha}_{{\bm q} \nu}$ generally don't cancel each other out, as in the one-dimensional model.  Furthermore, we find that within the plane the phonon angular momentum is dominantly pointing along the crystallographic directions, which is reminiscent of what we found in BaTiO$_3$ from the first principles (as shown in Fig.~\ref{fig:dist_planes_tet}). More details are provided in the supplement.\cite{supplement} 

Therefore, we conclude that the phonon anisotropy in our qualitative model is consistent with the anisotropy we found from first-principles.  

\section{Rhombohedral phase}\label{sec:rhomb}

\begin{figure}[t]
  \includegraphics{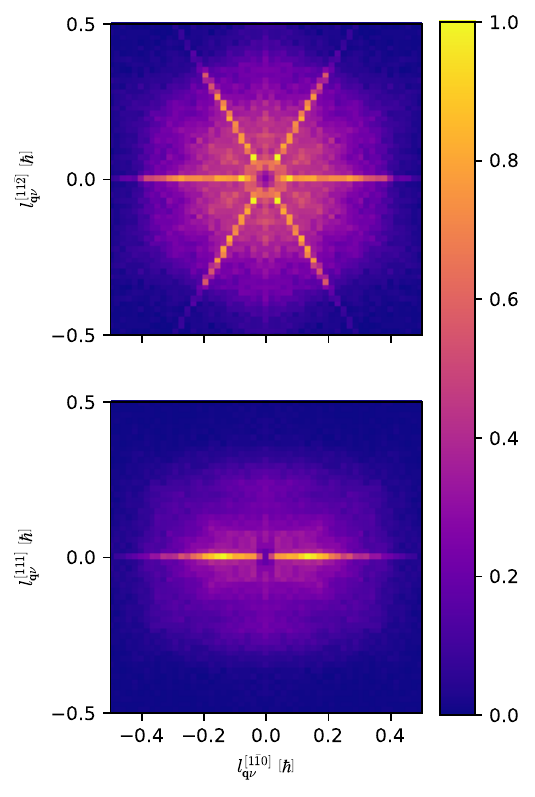}
  \caption{Same as Fig.~\ref{fig:dist_planes_tet}, but for rhombohedral phase of BaTiO$_3$. Top panel again shows distribution of phonon angular in plane perpendicular to $\bm P$, while bottom panel shows distribution in plane containing $\bm P$.} 
  \label{fig:dist_planes_rh}
\end{figure}
Now we analyze the phonon angular momentum in the rhombohedral phase of BaTiO$_3$. In this phase, $\bm P$ is pointing along the $[111]$ crystallographic direction. We define the averages
of $\bm l$ perpendicular to ${\bm P}$,
\begin{align}
\label{eq:LperpRho}
\big \langle l_{\perp}^{\rm rhom} \big \rangle
=
\frac{1}{N_q N_{\nu}} \sum_{\bm q} \sum_{\nu} \sqrt{
\left(
{\bm l}_{{\mathbf{q}}\nu} \cdot {\bm{\hat{n}}}_{1\bar{1}0}
\right)^2
+
\left(
{\bm l}_{{\mathbf{q}}\nu} \cdot {\bm{\hat{n}}}_{11\bar{2}}
\right)^2
}
\end{align}
and along $\bm P$,
\begin{align}
\label{eq:LparRho}
\big \langle l_{\parallel}^{\rm rhom} \big \rangle
=
\frac{1}{N_q N_{\nu}} \sum_{\bm q} \sum_{\nu} \left| {\bm l}_{{\mathbf{q}}\nu} \cdot {\bm{\hat{n}}}_{111} \right|.
\end{align}
Averages are once again performed over positive definite values, and the calculated values of the averaged angular momenta are
\begin{align}
\big \langle l_{\perp}^{\rm rhom} \big \rangle
& = 0.191\ \hbar, \notag \\
\big \langle l_{\parallel}^{\rm rhom} \big \rangle
& = 0.089 \ \hbar. \notag
\end{align}
As in the tetragonal phase of BaTiO$_3$, phonon angular momentum is larger perpendicular to the polarization than parallel. However, $\langle l_{\perp}^{\rm rhom} \rangle$ is only about 2 times larger than $\langle l_{\parallel}^{\rm rhom} \rangle$, so the anisotropy is significantly less than the tetragonal phase where the ratio was 6. The phonon angular momentum distributions which we show in Fig.~\ref{fig:dist_planes_rh} further illustrate this finding.  The bottom panel of Fig.~\ref{fig:dist_planes_rh} shows the distribution of phonon angular momentum in the plane spanned by $[1\bar{1}0]$ and $[111]$ directions.

The top panel of Fig.~\ref{fig:dist_planes_rh} shows the distributions of the angular momentum of the phonons in the plane perpendicular to $\bm P$.  This plane is spanned by crystallographic directions $[1\bar{1}0]$ and $[11\bar{2}]$. We find that only 32\% of the total phonon angular momentum is within $\pm 5^{\circ}$ of $[1\bar{1}0]$ and $[11\bar{2}]$, and their respective symmetry-related directions. This anisotropy is somewhat smaller than 47\% in-plane anisotropy in the tetragonal phase (see Fig.~\ref{fig:dist_planes_tet}).

Finally, we again linearly interpolated structures from the nonpolar to the polar rhombohedral phase. Once again, we observed phonon angular momentum saturation at $\lambda=0.5$, as in the tetragonal case.  Furthermore, the ratio of angular momentum perpendicular to parallel to $\bm P$ is consistently around 2. 

\section{Conclusion and suggested experiments}\label{sec:outlook}

The main result of this work is the calculated anisotropy of the angular momentum of the phonon relative to the electrical polarization $\bm P$ in BaTiO$_3$. The phonon angular momentum anisotropy is two-fold. First, the phonon angular momentum in the plane perpendicular to polarization $\bm P$ in the tetragonal phase is about six times higher than in the direction parallel to $\bm P$. Second, within the plane perpendicular to $\bm P$, about half of the phonon angular momentum is concentrated within $\pm 5^{\circ}$ of the high-symmetry crystal directions.

As these anisotropies in phonon angular momentum are tied to $\bm P$, it naturally follows that the reorientation of $\bm P$, induced by an external electric field, will then redistribute phonon angular momentum as well.  Therefore, any physical phenomenon, such as those listed in Sec.~\ref{sec:Intro}, that depends on the angular momentum of phonons in a ferroelectric, such as BaTiO$_3$, could be controlled by applying an external electric field.

Here, we focus on only one of the physical phenomena that rely on phonon angular momentum, the Einstein-de Haas effect.  This effect could be probed via ultrafast demagnetization experiments as in Ref.~\cite{nature_EdH}. As BaTiO$_3$ itself is not magnetic, one would need to couple BaTiO$_3$ to a magnetic material, consider a multiferroic material, or create a heterostructure between BaTiO$_3$ and a nonpolar magnetic perovskite. In the stacked geometry, the change in the angular momentum of the electron generated in the magnet has the opportunity to transfer into the phonon angular momentum in the adjacent BaTiO$_3$. Given our results (see Sec.~\ref{sec:results}), we predict that angular momentum transfer will depend on relative alignment (parallel or perpendicular) between magnetic domains ($\bm M$) and the BaTiO$_3$ polarization $\bm P$.

Finally, our work also resulted in a simple model that can give a qualitative understanding of the anisotropy in the angular momentum of phonons in BaTiO$_3$.  By studying a one-dimensional chain of Ti and O atoms, we find that polar displacements of atoms along the chain lead to no phonon angular momentum, within the assumptions of our model.  Nevertheless, polar displacements of atoms perpendicular to the chain lead to the phonon angular momentum that is perpendicular to both the chain and to the direction of atom displacements.  Our simple two-dimensional model, discussed in Sec.~\ref{sec:model2d} captures all qualitative features of the phonon angular momentum we found from the first principles in BaTiO$_3$.

\acknowledgments{This work was supported by the NSF DMR-1848074 grant. Computations were performed using the computer clusters and data storage resources of the HPCC at UCR, which were funded by grants from NSF (MRI-1429826) and NIH (1S10OD016290-01A1).}

\bibliography{pap}

\providecommand{\noopsort}[1]{}\providecommand{\singleletter}[1]{#1}%
\begin{thebibliography}{44}%
\makeatletter
\providecommand \@ifxundefined [1]{%
 \@ifx{#1\undefined}
}%
\providecommand \@ifnum [1]{%
 \ifnum #1\expandafter \@firstoftwo
 \else \expandafter \@secondoftwo
 \fi
}%
\providecommand \@ifx [1]{%
 \ifx #1\expandafter \@firstoftwo
 \else \expandafter \@secondoftwo
 \fi
}%
\providecommand \natexlab [1]{#1}%
\providecommand \enquote  [1]{``#1''}%
\providecommand \bibnamefont  [1]{#1}%
\providecommand \bibfnamefont [1]{#1}%
\providecommand \citenamefont [1]{#1}%
\providecommand \href@noop [0]{\@secondoftwo}%
\providecommand \href [0]{\begingroup \@sanitize@url \@href}%
\providecommand \@href[1]{\@@startlink{#1}\@@href}%
\providecommand \@@href[1]{\endgroup#1\@@endlink}%
\providecommand \@sanitize@url [0]{\catcode `\\12\catcode `\$12\catcode
  `\&12\catcode `\#12\catcode `\^12\catcode `\_12\catcode `\%12\relax}%
\providecommand \@@startlink[1]{}%
\providecommand \@@endlink[0]{}%
\providecommand \url  [0]{\begingroup\@sanitize@url \@url }%
\providecommand \@url [1]{\endgroup\@href {#1}{\urlprefix }}%
\providecommand \urlprefix  [0]{URL }%
\providecommand \Eprint [0]{\href }%
\providecommand \doibase [0]{https://doi.org/}%
\providecommand \selectlanguage [0]{\@gobble}%
\providecommand \bibinfo  [0]{\@secondoftwo}%
\providecommand \bibfield  [0]{\@secondoftwo}%
\providecommand \translation [1]{[#1]}%
\providecommand \BibitemOpen [0]{}%
\providecommand \bibitemStop [0]{}%
\providecommand \bibitemNoStop [0]{.\EOS\space}%
\providecommand \EOS [0]{\spacefactor3000\relax}%
\providecommand \BibitemShut  [1]{\csname bibitem#1\endcsname}%
\let\auto@bib@innerbib\@empty
\bibitem [{\citenamefont {McLellan}(1988)}]{McLellan}%
  \BibitemOpen
  \bibfield  {author} {\bibinfo {author} {\bibfnamefont {A.~G.}\ \bibnamefont
  {McLellan}},\ }\href
  {https://iopscience.iop.org/article/10.1088/0022-3719/21/7/009} {\bibfield
  {journal} {\bibinfo  {journal} {J. Phys. C: Solid State Phys.}\ }\textbf
  {\bibinfo {volume} {21}},\ \bibinfo {pages} {1177} (\bibinfo {year}
  {1988})}\BibitemShut {NoStop}%
\bibitem [{\citenamefont {Zhang}\ and\ \citenamefont {Niu}(2014)}]{Zhang2014}%
  \BibitemOpen
  \bibfield  {author} {\bibinfo {author} {\bibfnamefont {L.}~\bibnamefont
  {Zhang}}\ and\ \bibinfo {author} {\bibfnamefont {Q.}~\bibnamefont {Niu}},\
  }\href {https://doi.org/10.1103/PhysRevLett.112.085503} {\bibfield  {journal}
  {\bibinfo  {journal} {Phys. Rev. Lett.}\ }\textbf {\bibinfo {volume} {112}},\
  \bibinfo {pages} {085503} (\bibinfo {year} {2014})}\BibitemShut {NoStop}%
\bibitem [{\citenamefont {Streib}(2020)}]{streib2020difference}%
  \BibitemOpen
  \bibfield  {author} {\bibinfo {author} {\bibfnamefont {S.}~\bibnamefont
  {Streib}},\ }\bibfield  {title} {\bibinfo {title} {The difference between
  angular momentum and pseudo angular momentum},\ }\href@noop {} {\bibfield
  {journal} {\bibinfo  {journal} {arXiv preprint arXiv:2010.15616}\ } (\bibinfo
  {year} {2020})}\BibitemShut {NoStop}%
\bibitem [{\citenamefont {Zhu}\ \emph {et~al.}(2018)\citenamefont {Zhu},
  \citenamefont {Yi}, \citenamefont {Li}, \citenamefont {Xiao}, \citenamefont
  {Zhang}, \citenamefont {Yang}, \citenamefont {Kaind}, \citenamefont {Li},
  \citenamefont {Wang},\ and\ \citenamefont {Zhang}}]{Zhang2018}%
  \BibitemOpen
  \bibfield  {author} {\bibinfo {author} {\bibfnamefont {H.}~\bibnamefont
  {Zhu}}, \bibinfo {author} {\bibfnamefont {J.}~\bibnamefont {Yi}}, \bibinfo
  {author} {\bibfnamefont {M.-Y.}\ \bibnamefont {Li}}, \bibinfo {author}
  {\bibfnamefont {J.}~\bibnamefont {Xiao}}, \bibinfo {author} {\bibfnamefont
  {L.}~\bibnamefont {Zhang}}, \bibinfo {author} {\bibfnamefont {C.-W.}\
  \bibnamefont {Yang}}, \bibinfo {author} {\bibfnamefont {R.~A.}\ \bibnamefont
  {Kaind}}, \bibinfo {author} {\bibfnamefont {L.-J.}\ \bibnamefont {Li}},
  \bibinfo {author} {\bibfnamefont {Y.}~\bibnamefont {Wang}},\ and\ \bibinfo
  {author} {\bibfnamefont {X.}~\bibnamefont {Zhang}},\ }\href
  {https://science.sciencemag.org/content/359/6375/579} {\bibfield  {journal}
  {\bibinfo  {journal} {Science}\ }\textbf {\bibinfo {volume} {359}},\ \bibinfo
  {pages} {579} (\bibinfo {year} {2018})}\BibitemShut {NoStop}%
\bibitem [{\citenamefont {Li}\ \emph {et~al.}(2019)\citenamefont {Li},
  \citenamefont {Wang}, \citenamefont {Jin}, \citenamefont {Lu}, \citenamefont
  {Lian}, \citenamefont {Meng}, \citenamefont {Blei}, \citenamefont {Gao},
  \citenamefont {Taniguchi}, \citenamefont {Watanabe}, \citenamefont {Ren},
  \citenamefont {Cao}, \citenamefont {Tongay}, \citenamefont {Smirnov},
  \citenamefont {Zhang},\ and\ \citenamefont {Shi}}]{exciton_coupling}%
  \BibitemOpen
  \bibfield  {author} {\bibinfo {author} {\bibfnamefont {Z.}~\bibnamefont
  {Li}}, \bibinfo {author} {\bibfnamefont {T.}~\bibnamefont {Wang}}, \bibinfo
  {author} {\bibfnamefont {C.}~\bibnamefont {Jin}}, \bibinfo {author}
  {\bibfnamefont {Z.}~\bibnamefont {Lu}}, \bibinfo {author} {\bibfnamefont
  {Z.}~\bibnamefont {Lian}}, \bibinfo {author} {\bibfnamefont {Y.}~\bibnamefont
  {Meng}}, \bibinfo {author} {\bibfnamefont {M.}~\bibnamefont {Blei}}, \bibinfo
  {author} {\bibfnamefont {M.}~\bibnamefont {Gao}}, \bibinfo {author}
  {\bibfnamefont {T.}~\bibnamefont {Taniguchi}}, \bibinfo {author}
  {\bibfnamefont {K.}~\bibnamefont {Watanabe}}, \bibinfo {author}
  {\bibfnamefont {T.}~\bibnamefont {Ren}}, \bibinfo {author} {\bibfnamefont
  {T.}~\bibnamefont {Cao}}, \bibinfo {author} {\bibfnamefont {S.}~\bibnamefont
  {Tongay}}, \bibinfo {author} {\bibfnamefont {D.}~\bibnamefont {Smirnov}},
  \bibinfo {author} {\bibfnamefont {L.}~\bibnamefont {Zhang}},\ and\ \bibinfo
  {author} {\bibfnamefont {S.-F.}\ \bibnamefont {Shi}},\ }\bibfield  {title}
  {\bibinfo {title} {Momentum-dark intervalley exciton in monolayer tungsten
  diselenide brightened via chiral phonon},\ }\href@noop {} {\bibfield
  {journal} {\bibinfo  {journal} {ACS nano}\ }\textbf {\bibinfo {volume}
  {13}},\ \bibinfo {pages} {14107} (\bibinfo {year} {2019})}\BibitemShut
  {NoStop}%
\bibitem [{\citenamefont {Delhomme}\ \emph {et~al.}(2020)\citenamefont
  {Delhomme}, \citenamefont {Vaclavkova}, \citenamefont {Slobodeniuk},
  \citenamefont {Orlita}, \citenamefont {Potemski}, \citenamefont {Basko},
  \citenamefont {Watanabe}, \citenamefont {Taniguchi}, \citenamefont {Mauro},
  \citenamefont {Barreteau} \emph {et~al.}}]{exciton_coupling2}%
  \BibitemOpen
  \bibfield  {author} {\bibinfo {author} {\bibfnamefont {A.}~\bibnamefont
  {Delhomme}}, \bibinfo {author} {\bibfnamefont {D.}~\bibnamefont
  {Vaclavkova}}, \bibinfo {author} {\bibfnamefont {A.}~\bibnamefont
  {Slobodeniuk}}, \bibinfo {author} {\bibfnamefont {M.}~\bibnamefont {Orlita}},
  \bibinfo {author} {\bibfnamefont {M.}~\bibnamefont {Potemski}}, \bibinfo
  {author} {\bibfnamefont {D.}~\bibnamefont {Basko}}, \bibinfo {author}
  {\bibfnamefont {K.}~\bibnamefont {Watanabe}}, \bibinfo {author}
  {\bibfnamefont {T.}~\bibnamefont {Taniguchi}}, \bibinfo {author}
  {\bibfnamefont {D.}~\bibnamefont {Mauro}}, \bibinfo {author} {\bibfnamefont
  {C.}~\bibnamefont {Barreteau}}, \emph {et~al.},\ }\bibfield  {title}
  {\bibinfo {title} {Flipping exciton angular momentum with chiral phonons in
  mose2/wse2 heterobilayers},\ }\href@noop {} {\bibfield  {journal} {\bibinfo
  {journal} {2D Materials}\ }\textbf {\bibinfo {volume} {7}},\ \bibinfo {pages}
  {041002} (\bibinfo {year} {2020})}\BibitemShut {NoStop}%
\bibitem [{\citenamefont {Thingstad}\ \emph {et~al.}(2019)\citenamefont
  {Thingstad}, \citenamefont {Kamra}, \citenamefont {Brataas},\ and\
  \citenamefont {Sudb\o{}}}]{magnon_coupling}%
  \BibitemOpen
  \bibfield  {author} {\bibinfo {author} {\bibfnamefont {E.}~\bibnamefont
  {Thingstad}}, \bibinfo {author} {\bibfnamefont {A.}~\bibnamefont {Kamra}},
  \bibinfo {author} {\bibfnamefont {A.}~\bibnamefont {Brataas}},\ and\ \bibinfo
  {author} {\bibfnamefont {A.}~\bibnamefont {Sudb\o{}}},\ }\bibfield  {title}
  {\bibinfo {title} {Chiral phonon transport induced by topological magnons},\
  }\href {https://doi.org/10.1103/PhysRevLett.122.107201} {\bibfield  {journal}
  {\bibinfo  {journal} {Phys. Rev. Lett.}\ }\textbf {\bibinfo {volume} {122}},\
  \bibinfo {pages} {107201} (\bibinfo {year} {2019})}\BibitemShut {NoStop}%
\bibitem [{\citenamefont {Dornes}\ \emph {et~al.}(2019)\citenamefont {Dornes},
  \citenamefont {Acremann}, \citenamefont {Savoini}, \citenamefont {Kubli},
  \citenamefont {Neugebauer}, \citenamefont {Abreu}, \citenamefont {Huber},
  \citenamefont {Lantz}, \citenamefont {Vaz}, \citenamefont {Lemke},\ and\
  \citenamefont {et~al.}}]{nature_EdH}%
  \BibitemOpen
  \bibfield  {author} {\bibinfo {author} {\bibfnamefont {C.}~\bibnamefont
  {Dornes}}, \bibinfo {author} {\bibfnamefont {Y.}~\bibnamefont {Acremann}},
  \bibinfo {author} {\bibfnamefont {M.}~\bibnamefont {Savoini}}, \bibinfo
  {author} {\bibfnamefont {M.}~\bibnamefont {Kubli}}, \bibinfo {author}
  {\bibfnamefont {M.~J.}\ \bibnamefont {Neugebauer}}, \bibinfo {author}
  {\bibfnamefont {E.}~\bibnamefont {Abreu}}, \bibinfo {author} {\bibfnamefont
  {L.}~\bibnamefont {Huber}}, \bibinfo {author} {\bibfnamefont
  {G.}~\bibnamefont {Lantz}}, \bibinfo {author} {\bibfnamefont {C.~a.~F.}\
  \bibnamefont {Vaz}}, \bibinfo {author} {\bibfnamefont {H.}~\bibnamefont
  {Lemke}},\ and\ \bibinfo {author} {\bibnamefont {et~al.}},\ }\bibfield
  {title} {\bibinfo {title} {The ultrafast einstein–de haas effect},\ }\href
  {https://doi.org/10.1038/s41586-018-0822-7} {\bibfield  {journal} {\bibinfo
  {journal} {Nature}\ }\textbf {\bibinfo {volume} {565}},\ \bibinfo {pages}
  {209–212} (\bibinfo {year} {2019})}\BibitemShut {NoStop}%
\bibitem [{\citenamefont {Juraschek}\ and\ \citenamefont
  {Spaldin}(2019)}]{Spaldin2018}%
  \BibitemOpen
  \bibfield  {author} {\bibinfo {author} {\bibfnamefont {D.~M.}\ \bibnamefont
  {Juraschek}}\ and\ \bibinfo {author} {\bibfnamefont {N.~A.}\ \bibnamefont
  {Spaldin}},\ }\bibfield  {title} {\bibinfo {title} {Orbital magnetic moments
  of phonons},\ }\href {https://doi.org/10.1103/PhysRevMaterials.3.064405}
  {\bibfield  {journal} {\bibinfo  {journal} {Phys. Rev. Materials}\ }\textbf
  {\bibinfo {volume} {3}},\ \bibinfo {pages} {064405} (\bibinfo {year}
  {2019})}\BibitemShut {NoStop}%
\bibitem [{\citenamefont {Juraschek}\ \emph {et~al.}(2017)\citenamefont
  {Juraschek}, \citenamefont {Fechner}, \citenamefont {Balatsky},\ and\
  \citenamefont {Spaldin}}]{Spaldin2017}%
  \BibitemOpen
  \bibfield  {author} {\bibinfo {author} {\bibfnamefont {D.~M.}\ \bibnamefont
  {Juraschek}}, \bibinfo {author} {\bibfnamefont {M.}~\bibnamefont {Fechner}},
  \bibinfo {author} {\bibfnamefont {A.~V.}\ \bibnamefont {Balatsky}},\ and\
  \bibinfo {author} {\bibfnamefont {N.~A.}\ \bibnamefont {Spaldin}},\ }\href
  {https://doi.org/10.1103/PhysRevMaterials.1.014401} {\bibfield  {journal}
  {\bibinfo  {journal} {Phys. Rev. Materials}\ }\textbf {\bibinfo {volume}
  {1}},\ \bibinfo {pages} {014401} (\bibinfo {year} {2017})}\BibitemShut
  {NoStop}%
\bibitem [{\citenamefont {Park}\ and\ \citenamefont {Yang}(2020)}]{PAMHE_pap}%
  \BibitemOpen
  \bibfield  {author} {\bibinfo {author} {\bibfnamefont {S.}~\bibnamefont
  {Park}}\ and\ \bibinfo {author} {\bibfnamefont {B.-J.}\ \bibnamefont
  {Yang}},\ }\bibfield  {title} {\bibinfo {title} {Phonon angular momentum hall
  effect},\ }\href {https://doi.org/10.1021/acs.nanolett.0c03220} {\bibfield
  {journal} {\bibinfo  {journal} {Nano Letters}\ }\textbf {\bibinfo {volume}
  {20}},\ \bibinfo {pages} {7694} (\bibinfo {year} {2020})},\ \bibinfo {note}
  {pMID: 32955897},\ \Eprint
  {https://arxiv.org/abs/https://doi.org/10.1021/acs.nanolett.0c03220}
  {https://doi.org/10.1021/acs.nanolett.0c03220} \BibitemShut {NoStop}%
\bibitem [{\citenamefont {Strohm}\ \emph {et~al.}(2005)\citenamefont {Strohm},
  \citenamefont {Rikken},\ and\ \citenamefont {Wyder}}]{PHE_2005}%
  \BibitemOpen
  \bibfield  {author} {\bibinfo {author} {\bibfnamefont {C.}~\bibnamefont
  {Strohm}}, \bibinfo {author} {\bibfnamefont {G.~L. J.~A.}\ \bibnamefont
  {Rikken}},\ and\ \bibinfo {author} {\bibfnamefont {P.}~\bibnamefont
  {Wyder}},\ }\bibfield  {title} {\bibinfo {title} {Phenomenological evidence
  for the phonon hall effect},\ }\href
  {https://doi.org/10.1103/PhysRevLett.95.155901} {\bibfield  {journal}
  {\bibinfo  {journal} {Phys. Rev. Lett.}\ }\textbf {\bibinfo {volume} {95}},\
  \bibinfo {pages} {155901} (\bibinfo {year} {2005})}\BibitemShut {NoStop}%
\bibitem [{\citenamefont {Romao}(2019)}]{anomalous_therm_exp}%
  \BibitemOpen
  \bibfield  {author} {\bibinfo {author} {\bibfnamefont {C.~P.}\ \bibnamefont
  {Romao}},\ }\bibfield  {title} {\bibinfo {title} {Anomalous thermal expansion
  and chiral phonons in ${\mathrm{bib}}_{3}{\mathrm{o}}_{6}$},\ }\href
  {https://doi.org/10.1103/PhysRevB.100.060302} {\bibfield  {journal} {\bibinfo
   {journal} {Phys. Rev. B}\ }\textbf {\bibinfo {volume} {100}},\ \bibinfo
  {pages} {060302(R)} (\bibinfo {year} {2019})}\BibitemShut {NoStop}%
\bibitem [{\citenamefont {Grissonnanche}\ \emph {et~al.}(2020)\citenamefont
  {Grissonnanche}, \citenamefont {Thériault}, \citenamefont {Gourgout},
  \citenamefont {Boulanger}, \citenamefont {Lefrançois}, \citenamefont
  {Ataei}, \citenamefont {Laliberté}, \citenamefont {Dion}, \citenamefont
  {Zhou}, \citenamefont {Pyon},\ and\ \citenamefont
  {et~al.}}]{pseudogap_cuprates_2020}%
  \BibitemOpen
  \bibfield  {author} {\bibinfo {author} {\bibfnamefont {G.}~\bibnamefont
  {Grissonnanche}}, \bibinfo {author} {\bibfnamefont {S.}~\bibnamefont
  {Thériault}}, \bibinfo {author} {\bibfnamefont {A.}~\bibnamefont
  {Gourgout}}, \bibinfo {author} {\bibfnamefont {M.-E.}\ \bibnamefont
  {Boulanger}}, \bibinfo {author} {\bibfnamefont {E.}~\bibnamefont
  {Lefrançois}}, \bibinfo {author} {\bibfnamefont {A.}~\bibnamefont {Ataei}},
  \bibinfo {author} {\bibfnamefont {F.}~\bibnamefont {Laliberté}}, \bibinfo
  {author} {\bibfnamefont {M.}~\bibnamefont {Dion}}, \bibinfo {author}
  {\bibfnamefont {J.-S.}\ \bibnamefont {Zhou}}, \bibinfo {author}
  {\bibfnamefont {S.}~\bibnamefont {Pyon}},\ and\ \bibinfo {author}
  {\bibnamefont {et~al.}},\ }\bibfield  {title} {\bibinfo {title} {Chiral
  phonons in the pseudogap phase of cuprates},\ }\href
  {https://doi.org/10.1038/s41567-020-0965-y} {\bibfield  {journal} {\bibinfo
  {journal} {Nature Physics}\ }\textbf {\bibinfo {volume} {16}},\ \bibinfo
  {pages} {1108–1111} (\bibinfo {year} {2020})}\BibitemShut {NoStop}%
\bibitem [{\citenamefont {Hamada}\ \emph {et~al.}(2018)\citenamefont {Hamada},
  \citenamefont {Minamitani}, \citenamefont {Hirayama},\ and\ \citenamefont
  {Murakami}}]{Hamada2018}%
  \BibitemOpen
  \bibfield  {author} {\bibinfo {author} {\bibfnamefont {M.}~\bibnamefont
  {Hamada}}, \bibinfo {author} {\bibfnamefont {E.}~\bibnamefont {Minamitani}},
  \bibinfo {author} {\bibfnamefont {M.}~\bibnamefont {Hirayama}},\ and\
  \bibinfo {author} {\bibfnamefont {S.}~\bibnamefont {Murakami}},\ }\bibfield
  {title} {\bibinfo {title} {Phonon angular momentum induced by the temperature
  gradient},\ }\href {https://doi.org/10.1103/PhysRevLett.121.175301}
  {\bibfield  {journal} {\bibinfo  {journal} {Phys. Rev. Lett.}\ }\textbf
  {\bibinfo {volume} {121}},\ \bibinfo {pages} {175301} (\bibinfo {year}
  {2018})}\BibitemShut {NoStop}%
\bibitem [{\citenamefont {Hamada}(2021)}]{hamada_thesis}%
  \BibitemOpen
  \bibfield  {author} {\bibinfo {author} {\bibfnamefont {M.}~\bibnamefont
  {Hamada}},\ }\href@noop {} {\emph {\bibinfo {title} {Theory of generation and
  conversion of phonon angular momentum}}}\ (\bibinfo  {publisher} {Springer
  Nature},\ \bibinfo {year} {2021})\BibitemShut {NoStop}%
\bibitem [{\citenamefont {Hamada}\ and\ \citenamefont
  {Murakami}(2020)}]{hamada2020ph_rotoelectric}%
  \BibitemOpen
  \bibfield  {author} {\bibinfo {author} {\bibfnamefont {M.}~\bibnamefont
  {Hamada}}\ and\ \bibinfo {author} {\bibfnamefont {S.}~\bibnamefont
  {Murakami}},\ }\bibfield  {title} {\bibinfo {title} {Phonon rotoelectric
  effect},\ }\href {https://doi.org/10.1103/PhysRevB.101.144306} {\bibfield
  {journal} {\bibinfo  {journal} {Phys. Rev. B}\ }\textbf {\bibinfo {volume}
  {101}},\ \bibinfo {pages} {144306} (\bibinfo {year} {2020})}\BibitemShut
  {NoStop}%
\bibitem [{\citenamefont {Sonntag}\ \emph {et~al.}(2020)\citenamefont
  {Sonntag}, \citenamefont {Reichardt}, \citenamefont {Beschoten},\ and\
  \citenamefont {Stampfer}}]{graphene_ang_control}%
  \BibitemOpen
  \bibfield  {author} {\bibinfo {author} {\bibfnamefont {J.}~\bibnamefont
  {Sonntag}}, \bibinfo {author} {\bibfnamefont {S.}~\bibnamefont {Reichardt}},
  \bibinfo {author} {\bibfnamefont {B.}~\bibnamefont {Beschoten}},\ and\
  \bibinfo {author} {\bibfnamefont {C.}~\bibnamefont {Stampfer}},\ }\bibfield
  {title} {\bibinfo {title} {Electrical control over phonon polarization in
  strained graphene},\ }\href@noop {} {\bibfield  {journal} {\bibinfo
  {journal} {arXiv preprint arXiv:2012.11963}\ } (\bibinfo {year}
  {2020})}\BibitemShut {NoStop}%
\bibitem [{\citenamefont {Chen}\ \emph
  {et~al.}(2020{\natexlab{a}})\citenamefont {Chen}, \citenamefont {Kadic},\
  and\ \citenamefont {Wegener}}]{chiral_meta}%
  \BibitemOpen
  \bibfield  {author} {\bibinfo {author} {\bibfnamefont {Y.}~\bibnamefont
  {Chen}}, \bibinfo {author} {\bibfnamefont {M.}~\bibnamefont {Kadic}},\ and\
  \bibinfo {author} {\bibfnamefont {M.}~\bibnamefont {Wegener}},\ }\bibfield
  {title} {\bibinfo {title} {Chiral triclinic metamaterial crystals supporting
  isotropic acoustical activity and isotropic chiral phonons},\ }\href@noop {}
  {\bibfield  {journal} {\bibinfo  {journal} {arXiv preprint arXiv:2010.00410}\
  } (\bibinfo {year} {2020}{\natexlab{a}})}\BibitemShut {NoStop}%
\bibitem [{\citenamefont {Chen}\ \emph
  {et~al.}(2020{\natexlab{b}})\citenamefont {Chen}, \citenamefont {Kadic},
  \citenamefont {Guenneau},\ and\ \citenamefont {Wegener}}]{3D_quasicrystal}%
  \BibitemOpen
  \bibfield  {author} {\bibinfo {author} {\bibfnamefont {Y.}~\bibnamefont
  {Chen}}, \bibinfo {author} {\bibfnamefont {M.}~\bibnamefont {Kadic}},
  \bibinfo {author} {\bibfnamefont {S.}~\bibnamefont {Guenneau}},\ and\
  \bibinfo {author} {\bibfnamefont {M.}~\bibnamefont {Wegener}},\ }\bibfield
  {title} {\bibinfo {title} {Isotropic chiral acoustic phonons in 3d
  quasicrystalline metamaterials},\ }\href
  {https://doi.org/10.1103/PhysRevLett.124.235502} {\bibfield  {journal}
  {\bibinfo  {journal} {Phys. Rev. Lett.}\ }\textbf {\bibinfo {volume} {124}},\
  \bibinfo {pages} {235502} (\bibinfo {year} {2020}{\natexlab{b}})}\BibitemShut
  {NoStop}%
\bibitem [{\citenamefont {Chen}\ \emph {et~al.}(2017)\citenamefont {Chen},
  \citenamefont {Qin},\ and\ \citenamefont {Liu}}]{PAM_plasma}%
  \BibitemOpen
  \bibfield  {author} {\bibinfo {author} {\bibfnamefont {Q.}~\bibnamefont
  {Chen}}, \bibinfo {author} {\bibfnamefont {H.}~\bibnamefont {Qin}},\ and\
  \bibinfo {author} {\bibfnamefont {J.}~\bibnamefont {Liu}},\ }\bibfield
  {title} {\bibinfo {title} {Photons, phonons, and plasmons with orbital
  angular momentum in plasmas},\ }\href@noop {} {\bibfield  {journal} {\bibinfo
   {journal} {Scientific Reports}\ }\textbf {\bibinfo {volume} {7}},\ \bibinfo
  {pages} {41731} (\bibinfo {year} {2017})}\BibitemShut {NoStop}%
\bibitem [{\citenamefont {Coh}(2019)}]{classific_arxiv}%
  \BibitemOpen
  \bibfield  {author} {\bibinfo {author} {\bibfnamefont {S.}~\bibnamefont
  {Coh}},\ }\bibfield  {title} {\bibinfo {title} {Classification of materials
  with phonon angular momentum and microscopic origin of angular momentum},\
  }\href {https://arxiv.org/pdf/1911.05064.pdf} {\  (\bibinfo {year}
  {2019})}\BibitemShut {NoStop}%
\bibitem [{\citenamefont {Rajagopal}\ and\ \citenamefont
  {Srinivasan}(1962)}]{STO_model1962}%
  \BibitemOpen
  \bibfield  {author} {\bibinfo {author} {\bibfnamefont {A.}~\bibnamefont
  {Rajagopal}}\ and\ \bibinfo {author} {\bibfnamefont {R.}~\bibnamefont
  {Srinivasan}},\ }\bibfield  {title} {\bibinfo {title} {Lattice dynamics of
  cubic perovskite structures, in particular srtio3},\ }\href
  {https://doi.org/https://doi.org/10.1016/0022-3697(62)90523-1} {\bibfield
  {journal} {\bibinfo  {journal} {Journal of Physics and Chemistry of Solids}\
  }\textbf {\bibinfo {volume} {23}},\ \bibinfo {pages} {633 } (\bibinfo {year}
  {1962})}\BibitemShut {NoStop}%
\bibitem [{\citenamefont {Ghosez}\ \emph {et~al.}(1999)\citenamefont {Ghosez},
  \citenamefont {Cockayne}, \citenamefont {Waghmare},\ and\ \citenamefont
  {Rabe}}]{Ghosez1999}%
  \BibitemOpen
  \bibfield  {author} {\bibinfo {author} {\bibfnamefont {P.}~\bibnamefont
  {Ghosez}}, \bibinfo {author} {\bibfnamefont {E.}~\bibnamefont {Cockayne}},
  \bibinfo {author} {\bibfnamefont {U.~V.}\ \bibnamefont {Waghmare}},\ and\
  \bibinfo {author} {\bibfnamefont {K.~M.}\ \bibnamefont {Rabe}},\ }\bibfield
  {title} {\bibinfo {title} {Lattice dynamics of ${\mathrm{batio}}_{3},$
  ${\mathrm{pbtio}}_{3}$, and ${\mathrm{pbzro}}_{3}$: A comparative
  first-principles study},\ }\href {https://doi.org/10.1103/PhysRevB.60.836}
  {\bibfield  {journal} {\bibinfo  {journal} {Phys. Rev. B}\ }\textbf {\bibinfo
  {volume} {60}},\ \bibinfo {pages} {836} (\bibinfo {year} {1999})}\BibitemShut
  {NoStop}%
\bibitem [{\citenamefont {Tinte}\ \emph {et~al.}(1999)\citenamefont {Tinte},
  \citenamefont {Stachiotti}, \citenamefont {Sepliarsky}, \citenamefont
  {Migoni},\ and\ \citenamefont {Rodriguez}}]{Tinte_1999}%
  \BibitemOpen
  \bibfield  {author} {\bibinfo {author} {\bibfnamefont {S.}~\bibnamefont
  {Tinte}}, \bibinfo {author} {\bibfnamefont {M.~G.}\ \bibnamefont
  {Stachiotti}}, \bibinfo {author} {\bibfnamefont {M.}~\bibnamefont
  {Sepliarsky}}, \bibinfo {author} {\bibfnamefont {R.~L.}\ \bibnamefont
  {Migoni}},\ and\ \bibinfo {author} {\bibfnamefont {C.~O.}\ \bibnamefont
  {Rodriguez}},\ }\bibfield  {title} {\bibinfo {title} {Atomistic modelling of
  {BaTiO}3 based on first-principles calculations},\ }\href
  {https://doi.org/10.1088/0953-8984/11/48/325} {\bibfield  {journal} {\bibinfo
   {journal} {Journal of Physics: Condensed Matter}\ }\textbf {\bibinfo
  {volume} {11}},\ \bibinfo {pages} {9679} (\bibinfo {year}
  {1999})}\BibitemShut {NoStop}%
\bibitem [{\citenamefont {Seo}\ and\ \citenamefont {Ahn}(2013)}]{Seo2013}%
  \BibitemOpen
  \bibfield  {author} {\bibinfo {author} {\bibfnamefont {Y.-S.}\ \bibnamefont
  {Seo}}\ and\ \bibinfo {author} {\bibfnamefont {J.~S.}\ \bibnamefont {Ahn}},\
  }\href {https://doi.org/10.1103/PhysRevB.88.014114} {\bibfield  {journal}
  {\bibinfo  {journal} {Phys. Rev. B}\ }\textbf {\bibinfo {volume} {88}},\
  \bibinfo {pages} {014114} (\bibinfo {year} {2013})}\BibitemShut {NoStop}%
\bibitem [{\citenamefont {Giannozzi}\ \emph {et~al.}(2009)\citenamefont
  {Giannozzi}, \citenamefont {Baroni}, \citenamefont {Bonini}, \citenamefont
  {Calandra}, \citenamefont {Car}, \citenamefont {Cavazzoni}, \citenamefont
  {Ceresoli}, \citenamefont {Chiarotti}, \citenamefont {Cococcioni},
  \citenamefont {Dabo}, \citenamefont {Corso}, \citenamefont {de~Gironcoli},
  \citenamefont {Fabris}, \citenamefont {Fratesi}, \citenamefont {Gebauer},
  \citenamefont {Gerstmann}, \citenamefont {Gougoussis}, \citenamefont
  {Kokalj}, \citenamefont {Lazzeri}, \citenamefont {Martin-Samos},
  \citenamefont {Marzari}, \citenamefont {Mauri}, \citenamefont {Mazzarello},
  \citenamefont {Paolini}, \citenamefont {Pasquarello}, \citenamefont
  {Paulatto}, \citenamefont {Sbraccia}, \citenamefont {Scandolo}, \citenamefont
  {Sclauzero}, \citenamefont {Seitsonen}, \citenamefont {Smogunov},
  \citenamefont {Umari},\ and\ \citenamefont {Wentzcovitch}}]{Giannozzi_2009}%
  \BibitemOpen
  \bibfield  {author} {\bibinfo {author} {\bibfnamefont {P.}~\bibnamefont
  {Giannozzi}}, \bibinfo {author} {\bibfnamefont {S.}~\bibnamefont {Baroni}},
  \bibinfo {author} {\bibfnamefont {N.}~\bibnamefont {Bonini}}, \bibinfo
  {author} {\bibfnamefont {M.}~\bibnamefont {Calandra}}, \bibinfo {author}
  {\bibfnamefont {R.}~\bibnamefont {Car}}, \bibinfo {author} {\bibfnamefont
  {C.}~\bibnamefont {Cavazzoni}}, \bibinfo {author} {\bibfnamefont
  {D.}~\bibnamefont {Ceresoli}}, \bibinfo {author} {\bibfnamefont {G.~L.}\
  \bibnamefont {Chiarotti}}, \bibinfo {author} {\bibfnamefont {M.}~\bibnamefont
  {Cococcioni}}, \bibinfo {author} {\bibfnamefont {I.}~\bibnamefont {Dabo}},
  \bibinfo {author} {\bibfnamefont {A.~D.}\ \bibnamefont {Corso}}, \bibinfo
  {author} {\bibfnamefont {S.}~\bibnamefont {de~Gironcoli}}, \bibinfo {author}
  {\bibfnamefont {S.}~\bibnamefont {Fabris}}, \bibinfo {author} {\bibfnamefont
  {G.}~\bibnamefont {Fratesi}}, \bibinfo {author} {\bibfnamefont
  {R.}~\bibnamefont {Gebauer}}, \bibinfo {author} {\bibfnamefont
  {U.}~\bibnamefont {Gerstmann}}, \bibinfo {author} {\bibfnamefont
  {C.}~\bibnamefont {Gougoussis}}, \bibinfo {author} {\bibfnamefont
  {A.}~\bibnamefont {Kokalj}}, \bibinfo {author} {\bibfnamefont
  {M.}~\bibnamefont {Lazzeri}}, \bibinfo {author} {\bibfnamefont
  {L.}~\bibnamefont {Martin-Samos}}, \bibinfo {author} {\bibfnamefont
  {N.}~\bibnamefont {Marzari}}, \bibinfo {author} {\bibfnamefont
  {F.}~\bibnamefont {Mauri}}, \bibinfo {author} {\bibfnamefont
  {R.}~\bibnamefont {Mazzarello}}, \bibinfo {author} {\bibfnamefont
  {S.}~\bibnamefont {Paolini}}, \bibinfo {author} {\bibfnamefont
  {A.}~\bibnamefont {Pasquarello}}, \bibinfo {author} {\bibfnamefont
  {L.}~\bibnamefont {Paulatto}}, \bibinfo {author} {\bibfnamefont
  {C.}~\bibnamefont {Sbraccia}}, \bibinfo {author} {\bibfnamefont
  {S.}~\bibnamefont {Scandolo}}, \bibinfo {author} {\bibfnamefont
  {G.}~\bibnamefont {Sclauzero}}, \bibinfo {author} {\bibfnamefont {A.~P.}\
  \bibnamefont {Seitsonen}}, \bibinfo {author} {\bibfnamefont {A.}~\bibnamefont
  {Smogunov}}, \bibinfo {author} {\bibfnamefont {P.}~\bibnamefont {Umari}},\
  and\ \bibinfo {author} {\bibfnamefont {R.~M.}\ \bibnamefont {Wentzcovitch}},\
  }\bibfield  {title} {\bibinfo {title} {{QUANTUM} {ESPRESSO}: a modular and
  open-source software project for quantum simulations of materials},\ }\href
  {https://doi.org/10.1088/0953-8984/21/39/395502} {\bibfield  {journal}
  {\bibinfo  {journal} {Journal of Physics: Condensed Matter}\ }\textbf
  {\bibinfo {volume} {21}},\ \bibinfo {pages} {395502} (\bibinfo {year}
  {2009})}\BibitemShut {NoStop}%
\bibitem [{\citenamefont {Giannozzi}\ \emph {et~al.}(2017)\citenamefont
  {Giannozzi}, \citenamefont {Andreussi}, \citenamefont {Brumme}, \citenamefont
  {Bunau}, \citenamefont {Nardelli}, \citenamefont {Calandra}, \citenamefont
  {Car}, \citenamefont {Cavazzoni}, \citenamefont {Ceresoli}, \citenamefont
  {Cococcioni}, \citenamefont {Colonna}, \citenamefont {Carnimeo},
  \citenamefont {Corso}, \citenamefont {de~Gironcoli}, \citenamefont {Delugas},
  \citenamefont {DiStasio}, \citenamefont {Ferretti}, \citenamefont {Floris},
  \citenamefont {Fratesi}, \citenamefont {Fugallo}, \citenamefont {Gebauer},
  \citenamefont {Gerstmann}, \citenamefont {Giustino}, \citenamefont {Gorni},
  \citenamefont {Jia}, \citenamefont {Kawamura}, \citenamefont {Ko},
  \citenamefont {Kokalj}, \citenamefont {Kü{\c{c}}ükbenli}, \citenamefont
  {Lazzeri}, \citenamefont {Marsili}, \citenamefont {Marzari}, \citenamefont
  {Mauri}, \citenamefont {Nguyen}, \citenamefont {Nguyen}, \citenamefont {de-la
  Roza}, \citenamefont {Paulatto}, \citenamefont {Ponc{\'{e}}}, \citenamefont
  {Rocca}, \citenamefont {Sabatini}, \citenamefont {Santra}, \citenamefont
  {Schlipf}, \citenamefont {Seitsonen}, \citenamefont {Smogunov}, \citenamefont
  {Timrov}, \citenamefont {Thonhauser}, \citenamefont {Umari}, \citenamefont
  {Vast}, \citenamefont {Wu},\ and\ \citenamefont {Baroni}}]{QE_add_2017}%
  \BibitemOpen
  \bibfield  {author} {\bibinfo {author} {\bibfnamefont {P.}~\bibnamefont
  {Giannozzi}}, \bibinfo {author} {\bibfnamefont {O.}~\bibnamefont
  {Andreussi}}, \bibinfo {author} {\bibfnamefont {T.}~\bibnamefont {Brumme}},
  \bibinfo {author} {\bibfnamefont {O.}~\bibnamefont {Bunau}}, \bibinfo
  {author} {\bibfnamefont {M.~B.}\ \bibnamefont {Nardelli}}, \bibinfo {author}
  {\bibfnamefont {M.}~\bibnamefont {Calandra}}, \bibinfo {author}
  {\bibfnamefont {R.}~\bibnamefont {Car}}, \bibinfo {author} {\bibfnamefont
  {C.}~\bibnamefont {Cavazzoni}}, \bibinfo {author} {\bibfnamefont
  {D.}~\bibnamefont {Ceresoli}}, \bibinfo {author} {\bibfnamefont
  {M.}~\bibnamefont {Cococcioni}}, \bibinfo {author} {\bibfnamefont
  {N.}~\bibnamefont {Colonna}}, \bibinfo {author} {\bibfnamefont
  {I.}~\bibnamefont {Carnimeo}}, \bibinfo {author} {\bibfnamefont {A.~D.}\
  \bibnamefont {Corso}}, \bibinfo {author} {\bibfnamefont {S.}~\bibnamefont
  {de~Gironcoli}}, \bibinfo {author} {\bibfnamefont {P.}~\bibnamefont
  {Delugas}}, \bibinfo {author} {\bibfnamefont {R.~A.}\ \bibnamefont
  {DiStasio}}, \bibinfo {author} {\bibfnamefont {A.}~\bibnamefont {Ferretti}},
  \bibinfo {author} {\bibfnamefont {A.}~\bibnamefont {Floris}}, \bibinfo
  {author} {\bibfnamefont {G.}~\bibnamefont {Fratesi}}, \bibinfo {author}
  {\bibfnamefont {G.}~\bibnamefont {Fugallo}}, \bibinfo {author} {\bibfnamefont
  {R.}~\bibnamefont {Gebauer}}, \bibinfo {author} {\bibfnamefont
  {U.}~\bibnamefont {Gerstmann}}, \bibinfo {author} {\bibfnamefont
  {F.}~\bibnamefont {Giustino}}, \bibinfo {author} {\bibfnamefont
  {T.}~\bibnamefont {Gorni}}, \bibinfo {author} {\bibfnamefont
  {J.}~\bibnamefont {Jia}}, \bibinfo {author} {\bibfnamefont {M.}~\bibnamefont
  {Kawamura}}, \bibinfo {author} {\bibfnamefont {H.-Y.}\ \bibnamefont {Ko}},
  \bibinfo {author} {\bibfnamefont {A.}~\bibnamefont {Kokalj}}, \bibinfo
  {author} {\bibfnamefont {E.}~\bibnamefont {Kü{\c{c}}ükbenli}}, \bibinfo
  {author} {\bibfnamefont {M.}~\bibnamefont {Lazzeri}}, \bibinfo {author}
  {\bibfnamefont {M.}~\bibnamefont {Marsili}}, \bibinfo {author} {\bibfnamefont
  {N.}~\bibnamefont {Marzari}}, \bibinfo {author} {\bibfnamefont
  {F.}~\bibnamefont {Mauri}}, \bibinfo {author} {\bibfnamefont {N.~L.}\
  \bibnamefont {Nguyen}}, \bibinfo {author} {\bibfnamefont {H.-V.}\
  \bibnamefont {Nguyen}}, \bibinfo {author} {\bibfnamefont {A.~O.}\
  \bibnamefont {de-la Roza}}, \bibinfo {author} {\bibfnamefont
  {L.}~\bibnamefont {Paulatto}}, \bibinfo {author} {\bibfnamefont
  {S.}~\bibnamefont {Ponc{\'{e}}}}, \bibinfo {author} {\bibfnamefont
  {D.}~\bibnamefont {Rocca}}, \bibinfo {author} {\bibfnamefont
  {R.}~\bibnamefont {Sabatini}}, \bibinfo {author} {\bibfnamefont
  {B.}~\bibnamefont {Santra}}, \bibinfo {author} {\bibfnamefont
  {M.}~\bibnamefont {Schlipf}}, \bibinfo {author} {\bibfnamefont {A.~P.}\
  \bibnamefont {Seitsonen}}, \bibinfo {author} {\bibfnamefont {A.}~\bibnamefont
  {Smogunov}}, \bibinfo {author} {\bibfnamefont {I.}~\bibnamefont {Timrov}},
  \bibinfo {author} {\bibfnamefont {T.}~\bibnamefont {Thonhauser}}, \bibinfo
  {author} {\bibfnamefont {P.}~\bibnamefont {Umari}}, \bibinfo {author}
  {\bibfnamefont {N.}~\bibnamefont {Vast}}, \bibinfo {author} {\bibfnamefont
  {X.}~\bibnamefont {Wu}},\ and\ \bibinfo {author} {\bibfnamefont
  {S.}~\bibnamefont {Baroni}},\ }\bibfield  {title} {\bibinfo {title} {Advanced
  capabilities for materials modelling with quantum {ESPRESSO}},\ }\href
  {https://doi.org/10.1088/1361-648x/aa8f79} {\bibfield  {journal} {\bibinfo
  {journal} {Journal of Physics: Condensed Matter}\ }\textbf {\bibinfo {volume}
  {29}},\ \bibinfo {pages} {465901} (\bibinfo {year} {2017})}\BibitemShut
  {NoStop}%
\bibitem [{\citenamefont {Perdew}\ \emph {et~al.}(2008)\citenamefont {Perdew},
  \citenamefont {Ruzsinszky}, \citenamefont {Csonka}, \citenamefont {Vydrov},
  \citenamefont {Scuseria}, \citenamefont {Constantin}, \citenamefont {Zhou},\
  and\ \citenamefont {Burke}}]{PBEsol}%
  \BibitemOpen
  \bibfield  {author} {\bibinfo {author} {\bibfnamefont {J.~P.}\ \bibnamefont
  {Perdew}}, \bibinfo {author} {\bibfnamefont {A.}~\bibnamefont {Ruzsinszky}},
  \bibinfo {author} {\bibfnamefont {G.~I.}\ \bibnamefont {Csonka}}, \bibinfo
  {author} {\bibfnamefont {O.~A.}\ \bibnamefont {Vydrov}}, \bibinfo {author}
  {\bibfnamefont {G.~E.}\ \bibnamefont {Scuseria}}, \bibinfo {author}
  {\bibfnamefont {L.~A.}\ \bibnamefont {Constantin}}, \bibinfo {author}
  {\bibfnamefont {X.}~\bibnamefont {Zhou}},\ and\ \bibinfo {author}
  {\bibfnamefont {K.}~\bibnamefont {Burke}},\ }\bibfield  {title} {\bibinfo
  {title} {Restoring the density-gradient expansion for exchange in solids and
  surfaces},\ }\href {https://doi.org/10.1103/PhysRevLett.100.136406}
  {\bibfield  {journal} {\bibinfo  {journal} {Phys. Rev. Lett.}\ }\textbf
  {\bibinfo {volume} {100}},\ \bibinfo {pages} {136406} (\bibinfo {year}
  {2008})}\BibitemShut {NoStop}%
\bibitem [{\citenamefont {Garrity}\ \emph {et~al.}(2014)\citenamefont
  {Garrity}, \citenamefont {Bennett}, \citenamefont {Rabe},\ and\ \citenamefont
  {Vanderbilt}}]{GBRV}%
  \BibitemOpen
  \bibfield  {author} {\bibinfo {author} {\bibfnamefont {K.~F.}\ \bibnamefont
  {Garrity}}, \bibinfo {author} {\bibfnamefont {J.~W.}\ \bibnamefont
  {Bennett}}, \bibinfo {author} {\bibfnamefont {K.~M.}\ \bibnamefont {Rabe}},\
  and\ \bibinfo {author} {\bibfnamefont {D.}~\bibnamefont {Vanderbilt}},\
  }\href {https://www.physics.rutgers.edu/gbrv/} {\bibfield  {journal}
  {\bibinfo  {journal} {Comput.\ Mater.\ Sci.}\ }\textbf {\bibinfo {volume}
  {81}},\ \bibinfo {pages} {446} (\bibinfo {year} {2014})}\BibitemShut
  {NoStop}%
\bibitem [{\citenamefont {Yuk}\ \emph {et~al.}(2017)\citenamefont {Yuk},
  \citenamefont {Pitike}, \citenamefont {Nakhmanson}, \citenamefont
  {Eisenbach}, \citenamefont {Li},\ and\ \citenamefont {Cooper}}]{Cooper}%
  \BibitemOpen
  \bibfield  {author} {\bibinfo {author} {\bibfnamefont {S.~F.}\ \bibnamefont
  {Yuk}}, \bibinfo {author} {\bibfnamefont {K.~C.}\ \bibnamefont {Pitike}},
  \bibinfo {author} {\bibfnamefont {S.~M.}\ \bibnamefont {Nakhmanson}},
  \bibinfo {author} {\bibfnamefont {M.}~\bibnamefont {Eisenbach}}, \bibinfo
  {author} {\bibfnamefont {Y.~W.}\ \bibnamefont {Li}},\ and\ \bibinfo {author}
  {\bibfnamefont {V.~R.}\ \bibnamefont {Cooper}},\ }\href
  {https://www.nature.com/articles/srep43482} {\bibfield  {journal} {\bibinfo
  {journal} {Sci.\ Rep.}\ }\textbf {\bibinfo {volume} {7}},\ \bibinfo {pages}
  {43482} (\bibinfo {year} {2017})}\BibitemShut {NoStop}%
\bibitem [{\citenamefont {Baroni}\ \emph {et~al.}(2001)\citenamefont {Baroni},
  \citenamefont {de~Gironcoli}, \citenamefont {Dal~Corso},\ and\ \citenamefont
  {Giannozzi}}]{DFPT}%
  \BibitemOpen
  \bibfield  {author} {\bibinfo {author} {\bibfnamefont {S.}~\bibnamefont
  {Baroni}}, \bibinfo {author} {\bibfnamefont {S.}~\bibnamefont
  {de~Gironcoli}}, \bibinfo {author} {\bibfnamefont {A.}~\bibnamefont
  {Dal~Corso}},\ and\ \bibinfo {author} {\bibfnamefont {P.}~\bibnamefont
  {Giannozzi}},\ }\href {https://doi.org/10.1103/RevModPhys.73.515} {\bibfield
  {journal} {\bibinfo  {journal} {Rev. Mod. Phys.}\ }\textbf {\bibinfo {volume}
  {73}},\ \bibinfo {pages} {515} (\bibinfo {year} {2001})}\BibitemShut
  {NoStop}%
\bibitem [{\citenamefont {Momma}\ and\ \citenamefont {Izumi}(2008)}]{VESTA}%
  \BibitemOpen
  \bibfield  {author} {\bibinfo {author} {\bibfnamefont {K.}~\bibnamefont
  {Momma}}\ and\ \bibinfo {author} {\bibfnamefont {F.}~\bibnamefont {Izumi}},\
  }\bibfield  {title} {\bibinfo {title} {Vesta: a three-dimensional
  visualization system for electronic and structural analysis},\ }\href@noop {}
  {\bibfield  {journal} {\bibinfo  {journal} {Journal of Applied
  crystallography}\ }\textbf {\bibinfo {volume} {41}},\ \bibinfo {pages} {653}
  (\bibinfo {year} {2008})}\BibitemShut {NoStop}%
\bibitem [{\citenamefont {Zhang}\ and\ \citenamefont {Niu}(2015)}]{Zhang2015}%
  \BibitemOpen
  \bibfield  {author} {\bibinfo {author} {\bibfnamefont {L.}~\bibnamefont
  {Zhang}}\ and\ \bibinfo {author} {\bibfnamefont {Q.}~\bibnamefont {Niu}},\
  }\bibfield  {title} {\bibinfo {title} {Chiral phonons at high-symmetry points
  in monolayer hexagonal lattices},\ }\href
  {https://doi.org/10.1103/PhysRevLett.115.115502} {\bibfield  {journal}
  {\bibinfo  {journal} {Phys. Rev. Lett.}\ }\textbf {\bibinfo {volume} {115}},\
  \bibinfo {pages} {115502} (\bibinfo {year} {2015})}\BibitemShut {NoStop}%
\bibitem [{\citenamefont {Keeble}\ and\ \citenamefont
  {Thomas}(2009)}]{BTO_P4mm}%
  \BibitemOpen
  \bibfield  {author} {\bibinfo {author} {\bibfnamefont {D.~S.}\ \bibnamefont
  {Keeble}}\ and\ \bibinfo {author} {\bibfnamefont {P.~A.}\ \bibnamefont
  {Thomas}},\ }\bibfield  {title} {\bibinfo {title} {{On the tetragonality of
  the room-temperature ferroelectric phase of barium titanate, BaTiO${\sb
  3}$}},\ }\href {https://doi.org/10.1107/S0021889809008310} {\bibfield
  {journal} {\bibinfo  {journal} {Journal of Applied Crystallography}\ }\textbf
  {\bibinfo {volume} {42}},\ \bibinfo {pages} {480} (\bibinfo {year}
  {2009})}\BibitemShut {NoStop}%
\bibitem [{\citenamefont {Kwei}\ \emph {et~al.}(1993)\citenamefont {Kwei},
  \citenamefont {Lawson}, \citenamefont {Billinge},\ and\ \citenamefont
  {W.Cheong}}]{kwei}%
  \BibitemOpen
  \bibfield  {author} {\bibinfo {author} {\bibfnamefont {G.~H.}\ \bibnamefont
  {Kwei}}, \bibinfo {author} {\bibfnamefont {A.~C.}\ \bibnamefont {Lawson}},
  \bibinfo {author} {\bibfnamefont {S.~J.~L.}\ \bibnamefont {Billinge}},\ and\
  \bibinfo {author} {\bibfnamefont {S.}~\bibnamefont {W.Cheong}},\ }\href
  {https://pubs.acs.org/doi/pdf/10.1021/j100112a043} {\bibfield  {journal}
  {\bibinfo  {journal} {J.\ Phys.\ Chem.}\ }\textbf {\bibinfo {volume} {97}},\
  \bibinfo {pages} {2368} (\bibinfo {year} {1993})}\BibitemShut {NoStop}%
\bibitem [{Note1()}]{Note1}%
  \BibitemOpen
  \bibinfo {note} {If system has an inversion symmetry but the time-reversal
  symmetry is broken, then the phonon angular momentum at $\protect \bm {q}$
  and $-\protect \bm {q}$ has the same sign.}\BibitemShut {Stop}%
\bibitem [{Note2()}]{Note2}%
  \BibitemOpen
  \bibinfo {note} {For purposes of this example calculation, we used here
  structure of BaTiO$_3$ that is interpolated between cubic and tetragonal
  phase. More precisely, we set $\lambda ^{\protect \rm tet}$ parameter
  (defined in Sec.~\ref {sec:saturation}) to $0.2$.}\BibitemShut {Stop}%
\bibitem [{\citenamefont {Gonze}\ \emph {et~al.}(1994)\citenamefont {Gonze},
  \citenamefont {Charlier}, \citenamefont {Allan},\ and\ \citenamefont
  {Teter}}]{Gonze1994}%
  \BibitemOpen
  \bibfield  {author} {\bibinfo {author} {\bibfnamefont {X.}~\bibnamefont
  {Gonze}}, \bibinfo {author} {\bibfnamefont {J.}~\bibnamefont {Charlier}},
  \bibinfo {author} {\bibfnamefont {D.}~\bibnamefont {Allan}},\ and\ \bibinfo
  {author} {\bibfnamefont {M.}~\bibnamefont {Teter}},\ }\bibfield  {title}
  {\bibinfo {title} {Interatomic force constants from first principles: The
  case of \ensuremath{\alpha}-quartz},\ }\href
  {https://doi.org/10.1103/PhysRevB.50.13035} {\bibfield  {journal} {\bibinfo
  {journal} {Phys. Rev. B}\ }\textbf {\bibinfo {volume} {50}},\ \bibinfo
  {pages} {13035} (\bibinfo {year} {1994})}\BibitemShut {NoStop}%
\bibitem [{Note3()}]{Note3}%
  \BibitemOpen
  \bibinfo {note} {The remaining interatomic forces, not included in the sums
  above, are the onsite terms where both $i$ and $j{\protect \bm {R}}$
  correspond to the atoms in the home cell $\protect \bm {R}=0$. These on-site
  terms for both Ti and O sum to $0.54~\protect \rm {Ry/Bohr}^2$.}\BibitemShut
  {Stop}%
\bibitem [{\citenamefont {Kaxiras}(2003)}]{kaxiras}%
  \BibitemOpen
  \bibfield  {author} {\bibinfo {author} {\bibfnamefont {E.}~\bibnamefont
  {Kaxiras}},\ }\href@noop {} {\emph {\bibinfo {title} {Atomic and electronic
  structure of solids}}}\ (\bibinfo  {publisher} {Cambridge University Press},\
  \bibinfo {year} {2003})\BibitemShut {NoStop}%
\bibitem [{sup()}]{supplement}%
  \BibitemOpen
  \href@noop {} {}\bibinfo {howpublished} {URL will be inserted by
  publisher}\BibitemShut {NoStop}%
\bibitem [{Note4()}]{Note4}%
  \BibitemOpen
  \bibinfo {note} {It is clear from Eq.~\ref {eq:summed_ion_ang} that angular
  momentum $l_z=0$ unless both $x$ and $y$ components of phonon eigenvector
  $\xi $ are non-zero.}\BibitemShut {Stop}%
\bibitem [{Note5()}]{Note5}%
  \BibitemOpen
  \bibinfo {note} {With Fig.~\ref {fig:model}(a) in mind, one can select the
  center of any atom or spring as the origin, apply the inversion operator
  ($z\rightarrow -z$), and find the system unchanged.}\BibitemShut {Stop}%
\end{thebibliography}%

\foreach \x in {1,2}
{%
\clearpage


\section*{Supplementary material}

# Supplement for "Electric field control of phonon angular momentum in perovskite BaTiO$_3$"

Kevin Moseni,[1] Richard B. Wilson,[1,2] and Sinisa Coh[1,2]

[1]Materials Science and Engineering, University of California Riverside, Riverside, CA 92521, USA
[2]Mechanical Engineering, University of California Riverside, Riverside, CA 92521, USA

The python code provided with this supplement can be used to construct spring models used in the manuscript. The main function that creates a spring model is called `ph_model`. This function specifies parameters of the model such as number of periodic dimensions in the model (`dim_k`), number of dimensions of atomic displacements (`dim_d`), periodic lattice vectors (`lat`), Cartesian coordinates of atoms (`crt`), and masses of atoms (`mas`). Here is the syntax used to call this function,

```
ph_model(dim_k, dim_d, lat, crt, mas)
```

Once the model has been created with `ph_model` we use another function, called `set_spring` to specify the springs in the model. Here is the syntax of function `set_spring`,

```
set_spring(k_stretch, k_bending, i, j, R)
```

Here `k_stretch` is the $K_r$ parameter from the main paper specifying the stretching strength of the spring. Similarly `k_bending` specifies $K_\theta$ parameter. Finally, `i` specifies index of the first atom connected to this spring (atoms are indexed so that the first atom has index `0` not `1`), `j` specifies the second atom, and `R` specifies the unit cell vector of the second atom in terms of the periodic lattice vectors (`lat`). As periodicity is implicitly assumed in the code, it is enough to specify a spring between `i`-th and `j+R`-th atom, as this automatically specifies equivalent springs between `i+P`-th and `j+R+P`-th atom for all lattice vectors `P`. Similarly, by hermiticity specifying spring between `i`-th and `j+R`-th atom also specifies the same spring between `j`-th and `i-R`-th atom.

The following snippets of python code generate the models used in the main manuscript. The complete code that generates these models and creates all figures in this document are provided in the supplementary material.

In all cases we use the following, somewhat arbitrarily selected, numerical parameters in the code,

```
kr      = 1.0
krS     = 1.5
krL     = 1.0
ktheta  = 0.20
kthetaS = 0.24
kthetaL = 0.18
delta   = 0.1
m1      = 2.5
m2      = 1.0
```

## A. Model shown in Fig. 8 (b) in the main text

The model shown in Fig. 8 (b) consists of a chain of two atoms connected with a spring. One of the atoms is displaced perpendicular to the chain direction. This model can be specified with the following python code (see supplementary material for the entire code),

```
model = ph_model(dim_k = 1,
                 dim_d = 3,
                 lat = [[0.0, 0.0, 1.0]],
                 crt = [[0.0, delta*0.5, 0.0],
                        [0.0, 0.0       , 0.5]],
                 mas = [m1, m2])
model.set_spring(kr, ktheta, 0, 1, [ 0])
model.set_spring(kr, ktheta, 0, 1, [-1])
```

Below is the resulting phonon band structure.

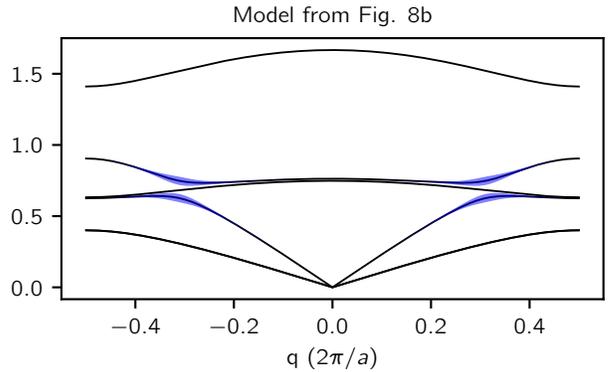

Blue color indicates phonon bands which have non-zero contribution of a single atom to the phonon angular momentum. Thickness of the blue curve is proportional to the contribution to the phonon angular momentum. As discussed in the main text, in this model total phonon angular momentum is zero, as contributions from two atoms in the unit cell cancel each other. This cancellation does not occur in the two-dimensional version of the model (as discussed shortly) or if we include springs between further neighboring atoms, as shown in the following figure.

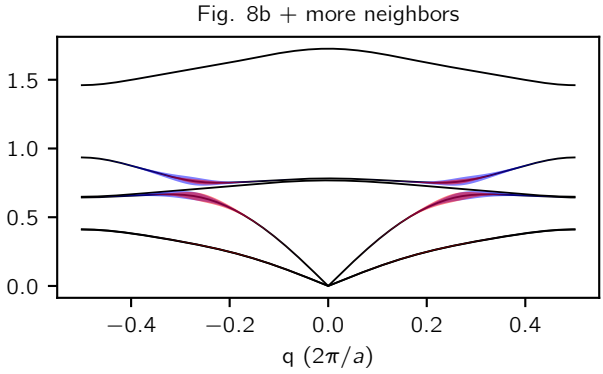

Fig. 8b + more neighbors

Here the red region is enhanced about 2000 times relative to the blue region.

### B. Model shown in Fig. 8 (c) in the main text

The model from Fig. 8 (c) can be generated by the following code,

```
model = ph_model(dim_k = 1,
                 dim_d = 3,
                 lat = [[0.0, 0.0, 1.0]],
                 crt = [[0.0, 0.0, delta*0.5],
                        [0.0, 0.0,       0.5]],
                 mas = [m1, m2])
model.set_spring(krS, kthetaS, 0, 1, [ 0])
model.set_spring(krL, kthetaL, 0, 1, [-1])
```

The resulting phonon band structure is,

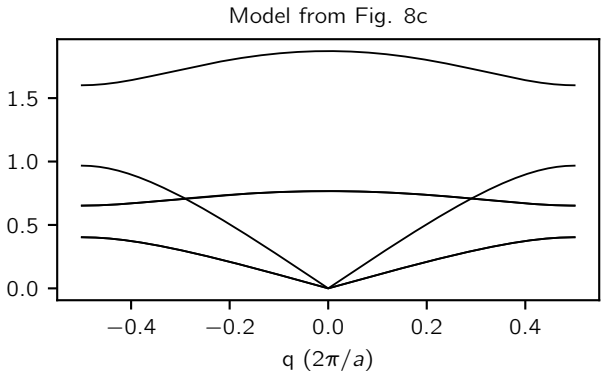

Model from Fig. 8c

As discussed in the main text, this model has both zero phonon angular momentum and zero contribution to the phonon angular momentum from a single atom in the unit cell. The same is true even if we include springs between further neighboring atoms.

### C. Two-dimensional model shown in Fig. 9 in the main text

Finally, the two-dimensional model from Fig. 9 in the main text can be generated with the following code,

```
model = ph_model(dim_k = 2,
                 dim_d = 3,
                 lat = [[1.0, 0.0, 0.0],
                        [0.0, 1.0, 0.0]],
                 crt = [[0.0, 0.0, delta*0.5],
                        [0.5, 0.0,       0.0],
                        [0.0, 0.5,       0.0]],
                 mas = [m1, m2, m2])
model.set_spring(kr, ktheta, 0, 1, [ 0, 0])
model.set_spring(kr, ktheta, 0, 2, [ 0, 0])
model.set_spring(kr, ktheta, 0, 1, [-1, 0])
model.set_spring(kr, ktheta, 0, 2, [ 0,-1])
```

The resulting phonon band structure shows again in blue contribution of individual atom to the phonon angular momentum. On the other hand, red color is used to show the total phonon angular momentum, when summed over all atoms in the unit cell. As discussed in the main text, in this model the total phonon angular momentum is non-zero. The trajectory in the reciprocal space of phonon wavevectors is a low-symmetry line that avoids all high-symmetry points in the Brillouin zone.

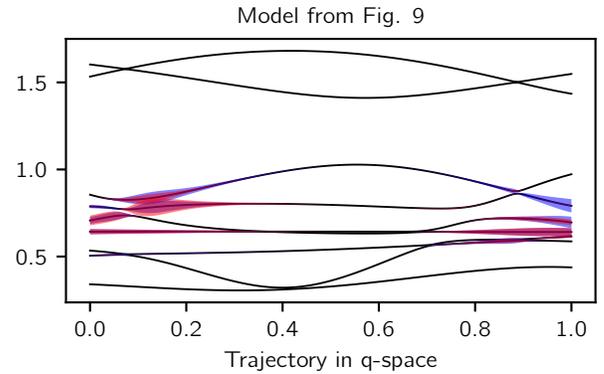

Model from Fig. 9

The two-dimensional histogram of the phonon angular moment for this model is shown below.

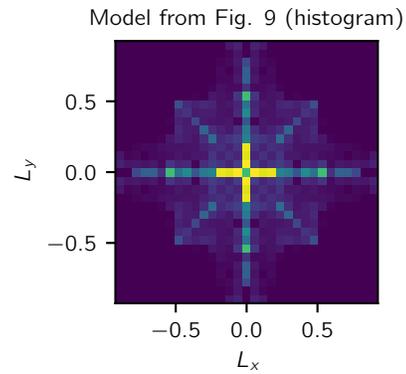

Model from Fig. 9 (histogram)

As can be seen from the figure, the qualitative features of the phonon angular momentum in this model are similar to that we found from first-principles, as shown in Fig. 2 in the main text.


}

\end{document}